\documentclass[12pt,draftcls,onecolumn]{IEEEtran}
\usepackage{amsbsy}
\usepackage{floatflt} 

\usepackage{amsmath}
\usepackage{amssymb}
\usepackage{times}
\usepackage{graphics}
\usepackage{graphicx}
\usepackage{xspace}
\usepackage{paralist} 
\usepackage{setspace} 
\usepackage{xypic}
\xyoption{curve}
\usepackage{latexsym}
\usepackage{theorem}
\usepackage{ifthen}
\usepackage{subfigure}
\usepackage{booktabs}
\usepackage{algorithm}
\usepackage{algorithmic}
\usepackage{color}

{\theoremheaderfont{\it} \theorembodyfont{\rmfamily}

}

{\theoremheaderfont{\it} \theorembodyfont{\rmfamily}
\newtheorem{theorem}{Theorem}
\newtheorem{lemma}[theorem]{Lemma}

\newtheorem{corollary}[theorem]{Corollary}

}

\title{Cooperative Slotted Aloha for Multi-Base Station Systems}

\author{Du$\check{\mbox{s}}$an Jakoveti\'c$^*$, Dragana Bajovi\'c, Dejan Vukobratovi\'c, and Vladimir Crnojevi\'c
\thanks{This paper was presented in part at the IEEE International Conference on Communications,
Workshop on Massive Uncoordinated Access Protocols, Sydney, Australia, June 2014;
in part at the European Wireless Conference, Barcelona, Spain, May 2014,
and in part at the IEEE International Symposium on Information Theory,
Honolulu, Hawaii, July 2014. The first and second authors are with University of Novi Sad, BioSense Center, Novi Sad, Serbia.
The third and fourth authors are with Department of Power, Electronics, and Communications Engineering, University of Novi Sad,
Novi Sad, Serbia. Authors' e-mails:
[djakovet, dbajovic, dejanv, crnojevic]@uns.ac.rs. $^*$corresponding author.}
\thanks{}}
\begin{document}
\maketitle \thispagestyle{empty} \maketitle
\vspace{-1.9cm}
\begin{abstract}
We introduce a framework to study slotted Aloha with cooperative base stations.
 Assuming a geographic-proximity communication model, we propose
several decoding algorithms
with different degrees of base stations' cooperation (non-cooperative, spatial, temporal, and spatio-temporal).
   With spatial cooperation, neighboring base stations inform each other
   whenever they collect a user within their coverage overlap; temporal cooperation
 corresponds to (temporal) successive interference cancellation done locally at each station.
  We analyze the four decoding algorithms and establish several fundamental results.
   With all algorithms, the peak throughput (average number of decoded users per slot, across all base stations)
 increases linearly with the number of base stations. Further, temporal and spatio-temporal cooperations exhibit
 a threshold behavior with respect to the normalized load (number of users per station, per slot).
 There exists a positive load $G^\star$, such that, below $G^\star$, the decoding probability
 is asymptotically maximal possible, equal the probability that a user is heard
 by at least one base station; with non-cooperative decoding and spatial cooperation, we show that $G^\star$ is zero.
   Finally, with spatio-temporal cooperation, we optimize the degree distribution according to
   which users transmit their packet replicas; the optimum is in general
   very different from the corresponding optimal distribution of the single-base station system.
\end{abstract}
\hspace{.43cm}\textbf{Keywords:} Slotted Aloha, successive interference cancellation, networked base stations, spatial cooperation, temporal cooperation, geometric random graphs.

\maketitle \thispagestyle{empty} \maketitle




%
\IEEEpeerreviewmaketitle


\vspace{-2mm}
\section{Introduction}
\label{section-intro}
We introduce a framework to study framed slotted Aloha with multiple, cooperative base stations.
We assume a geometric-proximity communication model, where users and base stations are
placed uniformly at random over a (unit) area, and the placements are mutually independent.
At each frame, each user transmits its packet replicas at multiple slots, according to a degree distribution
$\Lambda$, and is heard by all base stations within distance~$r$ from it.
 We develop and analyze several decoding algorithms that employ different degrees of cooperation
 across base stations (and across slots), namely: 1) non-cooperative decoding, spatial cooperation, temporal
 cooperation, and spatio-temporal cooperation. \emph{Spatial cooperation} allows for interference cancellation
 across neighboring base stations and works as follows. When a base station decodes a user, say $U_i$,
 at a certain slot, it informs other base stations that cover $U_i$ about its packet and its ID; subsequently,
 each of these stations subtracts the interference contribution from $U_i$ from its signal,
 which may reveal a singleton signal and allow the decoding of an additional user. With
 \emph{temporal cooperation}, each base station performs
 successive interference cancellation (SIC) (see, e.g.,~\cite{SIC10}) locally, across different slots in the frame,
 as, e.g., in~\cite{SlottedALOHAwithIC,liva}. Namely, when a base station
 observes a singleton in a certain slot, it decodes the corresponding user, say $U_i$, and subtracts its interference
 contribution from other slots where $U_i$ was active, which may result in additional singleton slots (and additional
 collected users). With spatio-temporal cooperation, spatial and temporal cooperations are alternated over several decoding iterations.

We establish several fundamental results with the four decoding algorithms.
 First, we show that, with all schemes, the peak throughput (expected number of
 decoded users per slot, across all base stations) increases linearly in
 the number of base stations~$m$. Next, we establish with
 temporal and spatio-temporal cooperations that there exists a threshold $G^\star$
 on the normalized load $G$ (number of users per slot, per base station), below
 which the decoding probability asymptotically equals its maximal possible value--the probability
 that a user is heard by at least one base station. We characterize the threshold
 $G^\star$ in terms of the threshold $H^\star$ of the single-base station slotted Aloha with
 SIC~\cite{liva}, where users transmit according to the same temporal degree distribution~$\Lambda$.
 Namely, we show that $G^\star \geq \frac{1}{4} \frac{H^\star}{\delta}$, where $\delta $
  is the users' average spatial degree--the average number of base stations that hear it.
   Further, we show that, with non-cooperative decoding and spatial cooperation, the threshold
   $G^\star(\delta)$ is zero.\footnote{In this paper, our focus is on the decoding probability and throughput, as in, e.g.,~\cite{liva};
  a detailed study of other metrics like delay and stability, e.g.,~\cite{StabilityVerdu}, is not considered here.} Next, with spatio-temporal cooperation, we
   find closed-form expressions for the users' (variable nodes')
    and check nodes' degree distributions in the underlying decoding graph; based on the latter,
    we give an and-or-tree heuristic to evaluate the decoding probability.
    We optimize the users' temporal degree distribution $\Lambda$ to
   maximize the threshold~$G^\bullet$ that corresponds to the and-or-tree equations.
   The optimized $\Lambda^{\bullet}$ is dependent on~$\delta$ and is, for very small $\delta$'s (of order $0.1$),
   close to the single-base station optimal distribution in~\cite{liva};
   for larger $\delta$'s--in the range of practical interest--the optimized
   $\Lambda^{\bullet}$ is close or equal to the constant-degree-two distribution in~\cite{SlottedALOHAwithIC}.
%
%

Our framework is inspired by machine-to-machine (M2M) communications in upcoming mobile cellular networks (such as
long-term evolution--LTE and advanced LTE: LTE-A), where a massive amount of IP-enabled devices seek access to a randomly deployed small-cell network. The proposed spatial and/or temporal interference cancellation is compatible with the LTE architecture where the neighboring cells are mutually inter-connected (see, e.g., X2 interface in LTE/LTE-A~\cite{LTEadvanced}). Upcoming trends such as Cloud Radio Access Networks (C-RAN) are also compatible with our proposal.

%

We now review the literature to help us further contrast our work from the existing work.
 Slotted Aloha has been proposed in the 70s,~\cite{Abramson}. With (framed) slotted Aloha~\cite{Fadra}, at each frame,
 each user transmits in one randomly selected slot. Reference~\cite{DiversityAloha} proposes a protocol
 where each user transmits in two randomly selected slots per frame.
  Reference~\cite{GeneralizedAloha} proposes a generalized slotted
      Aloha protocol where each user can be in two possible
      states, depending on whether its last packet transmission was decoded or not.
      Each user transmits in the next slot with a certain probability that depends on its current state.
       The paper obtains throughput bounds for cooperative users and explores the trade-off between throughput
and short-term fairness. 
   Reference~\cite{SlottedALOHAwithIC}
  significantly increases the achievable throughput with respect to standard slotted Aloha
    by incorporating the SIC mechanism into the protocol. Reference~\cite{liva} (see also~\cite{CedaNew,LivaInfoTheory})
    demonstrates that the protocol in~\cite{SlottedALOHAwithIC} is equivalent to
    the graph-peeling decoding of LDPC (low density parity check) codes over
    erasure channel (see, e.g.,~\cite{LDPC}) and exploits this analogy to improve the throughput.
     In~\cite{Gaudenzi4},
     the authors propose a spread-spectrum based random access
     with packet-oriented window memory-based SIC.
      Reference~\cite{Kissling1} proposes and analyzes an un-slotted Aloha protocol with SIC
      and shows its high performance in terms of
      packet loss ratio~(PLR) and throughput.
      Reference~\cite{Kissling2}
      further enhances~\cite{Kissling1} by incorporating a mechanism
      to resolve partial packet collisions. In~\cite{Gaudenzi2},
      the authors propose and analyze a novel asynchronous
      evolution of the scheme in~\cite{SlottedALOHAwithIC}; the
      scheme improves over~\cite{SlottedALOHAwithIC}, and, differently from~\cite{Kissling1,Kissling2},
      it operates asynchronously at the frame level as well.
    References~\cite{FramelessALOHA,CedomirAlohaRateless} achieve high throughputs
    via the frameless Aloha protocol by exploiting the analogy with rateless codes,
    while~\cite{CedaCapture1} analyzes frameless Aloha with capture effect.
     Reference~\cite{Herrero2}
     further enhances the protocol in~\cite{SlottedALOHAwithIC} by utilizing 3-5
     packet replica transmissions, and by exploiting power unbalance and capture.
     Recently, in~\cite{Gaudenzi1},
     the authors give a comprehensive analytical framework for slotted
     random access with and without SIC; the framework accounts for capture effect and
     accurately predicts random access performance--both in terms of PLR and throughput. Finally,
    \cite{CedaCapture2} considers Aloha with SIC and compressed sensing-based multi-user detection at the physical layer.
    Current paper is related to the above works in that it incorporates the SIC into
    random access protocols, but it differs from them by considering multiple, cooperative base stations (as opposed
    to the single base station systems in~\cite{SlottedALOHAwithIC,liva,FramelessALOHA,CedomirAlohaRateless,CedaCapture1,CedaCapture2,Gaudenzi1,Gaudenzi2,Gaudenzi4}.)
    %
     %
      %




Random access schemes with multiple receivers (or base stations) have been studied, e.g., in~\cite{ZorziSpatialDiversity,LivaNovo,Ephremides}. Reference~\cite{ZorziSpatialDiversity} studies the capture effect with multiple antennas in the presence of fading and shadowing. Reference~\cite{LivaNovo} assumes independent on-off fading across different user-receiver pairs and derives analytically the decoding probability, when each receiver works in isolation from other receivers. Our work is different from the above works, as it considers a different, geometric
communication model, and also incorporates inter-base station cooperation.
 Reference~\cite{Ephremides}
  considers multi-receiver, non-adaptive, slotted Aloha; they assume a geographic-proximity model that resembles
  ours. A
  difference from our paper is that~\cite{Ephremides} does not consider spatial nor temporal cooperations.
   Closest to this paper is reference~\cite{Gaudenzi3} which presents simulated system performance of the scheme proposed in~\cite{Gaudenzi4}
    in a realistic, S-band,
mobile satellite multi-beam scenario. The authors introduce, independently
 of our work~\cite{MASSAP2,ISITarxiv,EWpaper}, an inter-receiver (inter-gateway) SIC, as we do here. However,
they are not concerned with providing any analytical results.
  Finally, with respect to our work~\cite{MASSAP2,ISITarxiv,EWpaper}, current paper contributes with several
 new results, including optimization of the users' temporal degree distributions,
comparison with single-base station degree distributions proposed in the literature, e.g.,~\cite{SlottedALOHAwithIC,liva},
 and considerations of several physical layer aspects (See Section~\ref{section-discussion}).

It is worth noting that, generally, interference cancellation across different base stations has been previously
considered in the literature, in contexts different than random access, e.g., TDMA (time division multiple access) and CDMA
(code division multiple access), see, e.g.,~\cite{hanly2D,ref160Bavarian,HanlyTutorial}, and references therein.
  For example,~\cite{hanly2D} considers TDMA cellular systems and proposes a belief-propagation-type  decoding
 for a 2-dimensional Wyner model.
 With respect to the above works, our work contrasts by the following.
 While the literature usually assumes Wyner-type (grid) communication models,
 our model is a geometric random model. Consequently, the underlying decoding graphs are
 very different--grid graphs versus random geometric graphs. Further, we consider random access,
 while the other works usually consider TDMA or CDMA systems.

\textbf{Paper organization}. The next paragraph introduces notation. Section~\ref{section-model-prel}
explains the model that we assume and gives preliminaries needed for subsequent analysis. Section~\ref{sec-Algs}
presents our four decoding algorithms. In Section~\ref{section-performance-analysis}, we analyze the algorithms' performance.
Section~\ref{section-simulations} performs numerical optimization of the users' temporal degree
distribution with spatio-temporal cooperation and provides simulation studies.
 Section~\ref{section-discussion} includes a discussion about
 assumptions made in the paper and about physical layer issues.
Finally, we conclude in Section~\ref{section-conclusion}.
 The remaining
proofs can be found in the supplementary material.

%

\textbf{Notation}. We denote by: ${\mathbb R}^d$ the $d$-dimensional Euclidean space;
$v_i$ the $i$-the entry of a vector $v$;
$\mathbf{B}(q,s)=$ $\{x \in {\mathbb R}^2:$$\,(x_1-q_1)^2+(x_2-q_2)^2 \leq s^2 \}$
the Euclidean ball in ${\mathbb R}^2$ centered at $q$ with radius $s$;
$\mathbf{B}_{\infty}(q,s)= $ $\{x \in{\mathbb R}^2:$$\,|x_1-q_1| \leq s,\,|x_2-q_2|\leq s \}$ the square centered at $q$, with the side length equal to~$2 s$; $\mathbf{R}(q,s_1,s_2)=$ $\{x \in {\mathbb R}^2:$$\,(x_1-q_1)^2+(x_2-q_2)^2 \in [s_1^2,s_2^2] \}$ the ring centered at $q$ with inner radius $s_1$ and outer radius $s_2$; $\mathcal{S}_1 \setminus \mathcal{S}_2$ the set difference between the sets $\mathcal{S}_1$ and $\mathcal{S}_2$; $|\mathcal{S}|$ the cardinality of set $\mathcal{S}$;  $1_{E}$ the indicator of event $E$; $\mathbb P$, $\mathbb E$, and $\mathrm{Var}$ the probability, expectation, and variance operators, respectively;
and ${\imath}$ the imaginary unit.

\vspace{-5mm}
\section{Model and preliminaries}
\label{section-model-prel}
This section introduces the system model that we assume
and gives preliminaries needed for the presentation of our algorithms and results.
Subsection~\ref{subsection-model} explains the model, while
Subsection~\ref{subsection-single-base-stations} reviews
single-base station slotted Aloha with and without (temporal) SIC.
 Finally, Subsection~\ref{subsection-perf-metrics} introduces performance metrics
that we study.
%
%
%
\vspace{-3mm}
\subsection{System model}
\label{subsection-model}
We consider framed slotted Aloha with $n$ users, $m$ base stations, and $\tau$ slots per frame.
  (The number of users $n$ is fixed.)
  Let $U_i$ denote user $i$, $i=1,...,n$, and $B_l$ base station $l$, $l=1,...,m$. The normalized load $G=n/(\tau m)$
  equals the number of users per base station, per slot. We assume that
  base stations are synchronized, in the sense that their slots
  are aligned in time, have equal duration, and there is an equal number of slots (equal $\tau$) at each base station.
  Henceforth, there are $t=1,...,\tau$ system-wide slots, at each frame.

\begin{figure}[thpb]
      \centering
      \vspace{-5mm}
       \includegraphics[height=2.7 in,width=3.2 in]{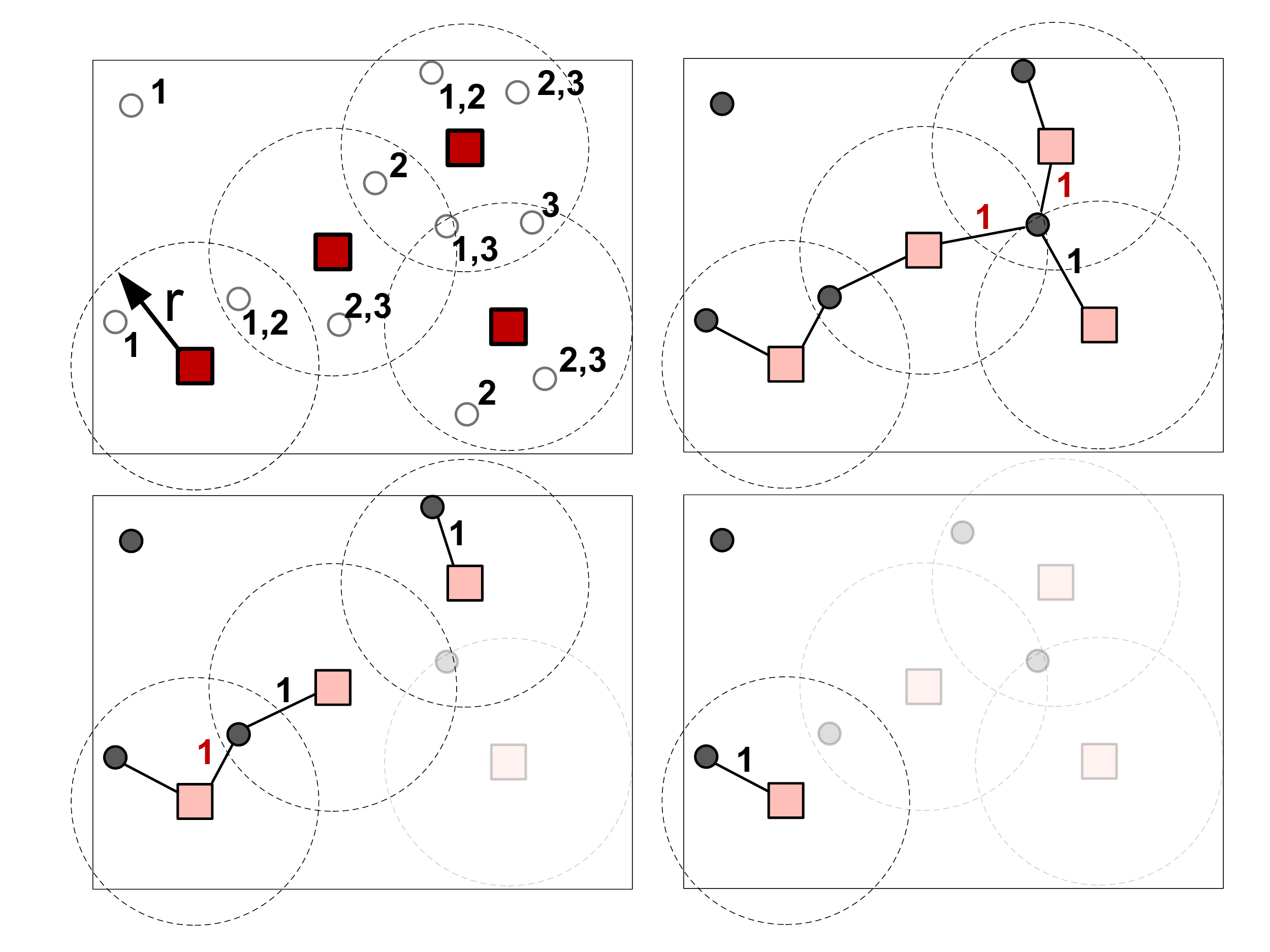}
       \includegraphics[height=2.65 in,width=3.1 in]{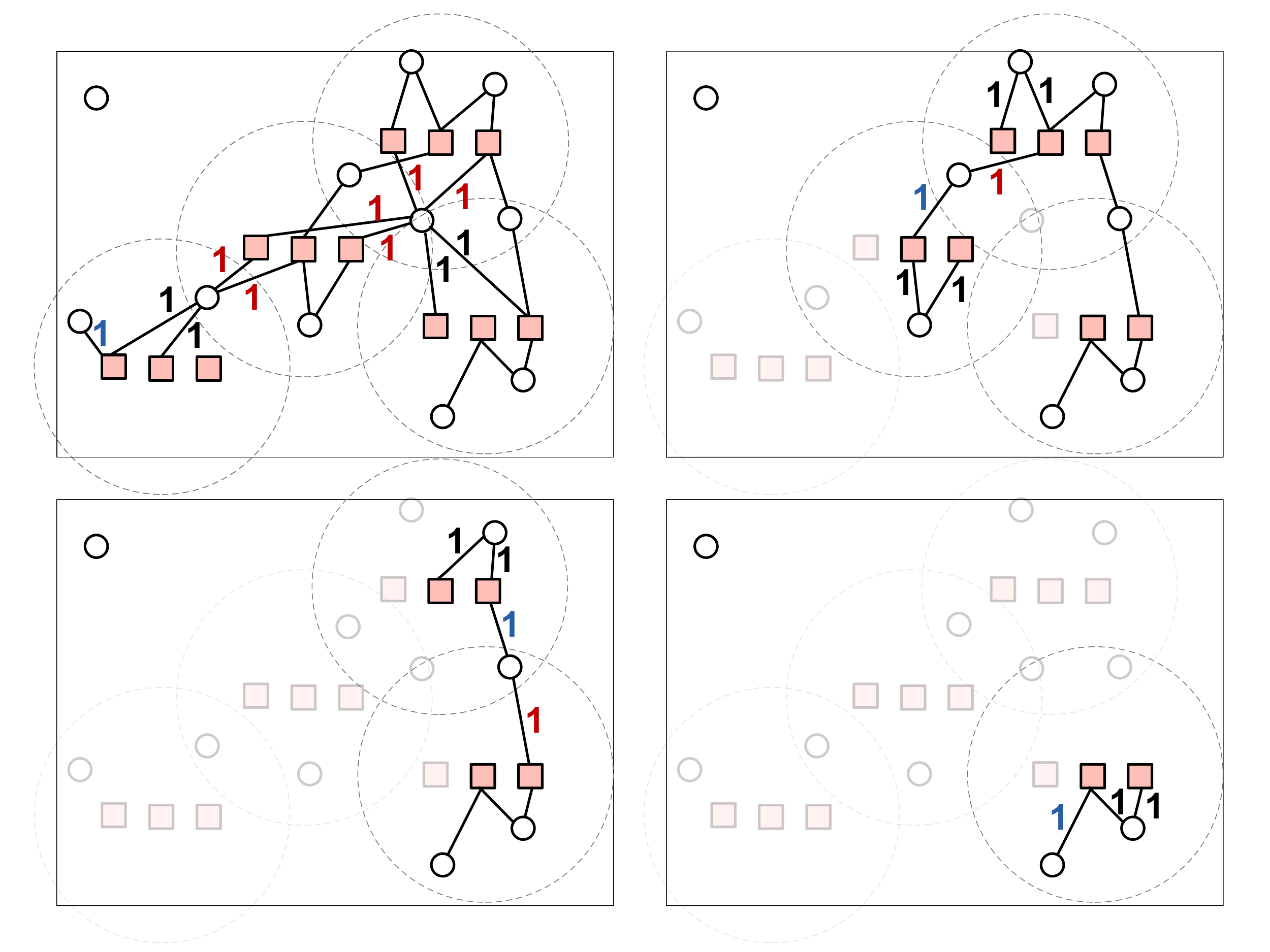}
       \vspace{-5mm}
       \caption{\emph{The group of four figures on the left}: System example with $m=4$ base stations, $n=11$ users, and $\tau=3$ slots (top left).
       Base stations are represented as red or pink squares,
        and users are represented as circles. The users' activation slots are indicated by numbers next to each user. The three figures (top right and two bottom figures)
       give an example of spatial cooperation decoding at slot~$t=1$.  Top right: initial graph $\mathcal G_0$, introduced in Section~\ref{sec-Algs}, for slot~$t=1$. Symbols ``1'''s represent decoded links. A link is decoded at iteration~$s$ if it is adjacent to a user collected at~$s$. Black ``1'''s are the links that are decoded locally, while red ``1'''s are the links revealed through communication among base stations. The sequence of figures top right, bottom left, bottom right represents decoding iterations $s=1,2,3$.
       \emph{The group of four figures on the right}: Spatio-temporal cooperation decoding for the
       depicted system example. Top left: initial graph $\mathcal H_0$, introduced in Section~\ref{sec-Algs}.
       Each base station has $\tau=3$ check nodes (pink squares),
       that correspond to three different slots (from left to right).
       The sequence of figures top left, top right, bottom left, and bottom right shows decoding iterations $s=1,2,3,4$. Black and red ``1'''s have the same meaning as with spatial cooperation, while blue ``1''s are the links decoded locally through temporal~SIC.}
       \label{figure-primer-1}
       \vspace{-5mm}
\end{figure}

\textbf{Transmission protocol and communication model}. At each frame, each user $U_i$
   transmits several replicas of the same message; each $U_i$'s message contains
   its information packet, its unique ID, and the pointer to all the slots at which $U_i$ transmits
    in a given frame.\footnote{With non-cooperative decoding and spatial cooperation,
    the pointer to the slots where $U_i$ is active is not needed and hence is not included in the message.} If $U_i$ transmits at a certain slot~$t$, we say that it is active at~$t$.
    Different users transmit mutually independently, each transmitting according to
    a degree distribution $\Lambda=(\Lambda_1,...,\Lambda_{s_{\mathrm{max}}})^\top$, $s_{\mathrm{max}} \leq \tau$. Here, $\Lambda_s
    = \mathbb P (Q_i=s)$, where $Q_i$ is the users' temporal degree, i.e., the number of
    slots per frame at which $U_i$ transmits.
     User $U_i$ transmits as follows. It generates a sample $Q_i$ from distribution ${\Lambda}$;
     if $Q_i=s$, then $U_i$ transmits in $s$ uniformly randomly selected slots. Denote by
     $\lambda:=\mathbb E [Q_i]=\sum_{s=1}^{s_{\mathrm{max}}} s \Lambda_s$ the users' average temporal degree. We assume that, whenever
     $U_i$ transmits, it is heard by all base stations within distance $r$ from it; likewise,
     each station $B_l$ hears a superposition of the signals of all active users within distance $r$ from it.
     (See Figure~\ref{figure-primer-1}, the left four figures--top left, for a system illustration.)
      If $U_i$ and $B_l$ are within distance~$r$, we say they are adjacent.

\textbf{Placement model}.
 All users and base stations
are placed over a unit square~$\mathcal A:=\mathbf{B}_{\infty}(0,1/2)$.\footnote{All our results
hold unchanged (except Theorem~1~(a) which holds under a minor modification)
for the unit disk area, as well; we adopt the unit square as it is common with random geometric
graph-type models, e.g.,~\cite{BoydGossip}.}
%
 Each user $U_i$ is placed uniformly at random over~$\mathcal A$.
 We denote by $u_i \in \mathcal A$ the random placement of $U_i$.
 Each base station $B_l$ is positioned at a random location~$b_l$, generated uniformly
 at random over $\mathcal A$. All the placements, $u_i$, $i=1,...,n$, $b_l$, $l=1,...,m$,
 are mutually independent, and they are fixed during each frame. We distinguish two types of
 users' and base stations' placements: 1) nominal placements, that fall within $\mathcal A^{\mathrm{o},r}:=\mathbf{B}_{\infty}(0,1/2-2 r)$;
 and 2) boundary placements, within $\partial A := \mathcal A \setminus \mathcal A^{\mathrm{o},r}$, $r \leq 1/4$.
 We let $\delta:=m r^2 \pi$. The quantity $\delta$ equals the average number of base stations that hear a
 nominally placed user. We refer to $\delta$ as the users' average spatial degree. (See also ahead Section~\ref{sec-Algs}
 for the graph representation of the system.) We present our decoding algorithms in Section~\ref{sec-Algs}. Throughout the paper, we assume
that a user $U_i$ is decoded if it is decoded by at least one adjacent base station; if the latter occurs,
we say that $U_i$ is collected by the system. For a fixed user $U_i$, we denote by $\mathbb P(U_i\,\mathrm{coll.})$
 the probability $U_i$ is collected.
 Note that $1-\mathbb P(U_i\,\mathrm{coll.})$ equals the packet loss ratio~(PLR); see, e.g.,~\cite{SlottedALOHAwithIC,Kissling1,liva}.

\vspace{-3mm}
\subsection{Single base station systems}
\label{subsection-single-base-stations}
One of our goals is to examine the throughput gains of each decoding algorithm when multiple ($m$) base stations are introduced,
as opposed to standard single-base station systems. Hence, for future comparisons,
we briefly describe two standard single base station systems: 1) slotted Aloha; and 2) slotted Aloha with
(temporal) SIC,~\cite{liva}. With both systems, the time slots are framed, the base station is placed at the center of the region,
and its radius~$r$ is large enough to cover all users. For both systems, we let $H$ be the load--total number of users
divided by the total number of slots within each frame. With slotted Aloha, each user transmits its message (containing its information packet) in one uniformly
randomly selected slot within the frame.
  Base station decodes a user at a certain slot if and only if it observes a singleton (exactly one user transmitted at the slot).
 Asymptotically,\footnote{The asymptotic setting is such that the number of users and
the number of slots both grow to infinity, but their ratio (load) converges to a positive constant~$H$.}
the decoding probability $\mathbb P(U_i\,\mathrm{coll.})$ is~$\mathrm{exp}(-H)$, the throughput (expected number of collected users per slot) is $H\mathrm{exp}(-H)$, and the peak throughput is $1/e$--achieved at $H=1$.

Regarding slotted Aloha with temporal SIC~\cite{liva}, users transmit their messages in multiple slots according to a distribution~$\Lambda$, and each user
transmits independently from other users. Each message of each user
contains the information packet and the list of all slots where the user transmits. After all transmissions within the frame are completed,
the base station performs an iterative decoding as follows. At iteration~$s$, it checks whether there are any singleton slots.
 If there are singleton slots, the base station selects one of them, say slot~$t$, collects a user, say~$U_i$, and recovers the $U_i$'s
 list of its remaining activation slots. Subsequently, the base station subtracts the interference contribution of $U_i$
  in each remaining $U_i$'s activation slot.\footnote{More precisely, base station reconstructs the waveform
  that corresponds to the $U_i$'s information packet and subtracts it from the signal waveforms
  that correspond to each remaining $U_i$'s activation slot.} Note that this operation may reveal additional singleton slots.
  Subsequently, the base station proceeds to the next iteration and looks for the singleton slots.
  The iterations continue until the base station observes no singleton slots. The decoding probability $\mathbb P(U_i\,\mathrm{coll.})$ with this scheme asymptotically exhibits
a threshold behavior. Denote by ${\rho}(H)$
 the asymptotic decoding probability at load~$H$.\footnote{The asymptotic setting is as follows.
 Fix the number of decoding iterations to~$s$, the number of nodes $n$, the number of slots $\tau=\tau(n)$, and $n=H \tau(n)$, $\forall n$.
 Then, ${\rho}(H)$ is defined as
 $\lim_{s \rightarrow \infty}\lim_{n \rightarrow \infty} \mathbb P(U_i\,\mathrm{coll.})$.} There exists a strictly positive load $H^\star$,
 defined as the largest load $H^\prime$ such that ${\rho}(H)=1$, $\forall H \leq H^\prime$. (This should be
contrasted with the standard slotted Aloha, where the decoding probability is $\mathrm{exp}(-H)$
 and is strictly below one for arbitrarily small~$H$.)
  The corresponding (asymptotic) peak throughput
   can be made arbitrarily close to $1$, see~\cite{Narayanan2012,liva2012spatially}.
 For arbitrary load~$H$, asymptotic values of
decoding probability and throughput are not given in closed form, but can be evaluated via and-or-tree formulas; see~\cite{liva} for the details.
\vspace{-3mm}
\subsection{Performance metrics}
\label{subsection-perf-metrics}

We will usually be interested in the asymptotic setting,
defined as follows. The number of: users~$n$, base stations~$m=m(n)$, and slots~$\tau=\tau(n)$ all converge to infinity, and the communication
radius~$r=r(n)$ goes to zero, such that the users' average spatial degree $m r^2 \pi \rightarrow \delta$, and
the normalized load $n/(\tau m) \rightarrow G$, where $\delta$ and $G$ are positive constants.
(We assume that, when $\tau\rightarrow \infty$, $s_{\mathrm{max}}$ in the users'
temporal degree distribution $\Lambda=(\Lambda_1,...,\Lambda_{s_{\mathrm{max}}})^\top$ remains finite.) Throughout,
when we state that a certain result holds asymptotically, it is in the sense of the above setting.

Denote by~$\mathbb P (U_i\,\mathrm{cov.})$ the probability that a user is covered by at least one base station.
Clearly, this is the probability that the $U_i$'s spatial degree is strictly greater than zero, and equals asymptotically $1-\mathrm{exp}(-\delta)$.\footnote{This is because the $U_i$'s spatial degree asymptotically
follows a Poisson distribution with parameter $\delta$; See ahead Section~{III}, paragraph with Heading
Degree distributions in $\mathcal G_0$.} Also, it is clear that, for any decoding algorithm, we must have $\mathbb P (U_i\,\mathrm{coll.}) \leq \mathbb P (U_i\,\mathrm{cov.})$.
 Throughout the paper, we restrict to the range of~$\delta$'s that ensure a prescribed $1-\epsilon$ coverage requirement, where
 $\epsilon>0$ is a small constant; that is, given a $1-\epsilon$ coverage requirement, we let~$\delta \geq \mathrm{ln}(1/\epsilon)$.

Expected fraction of collected users is given by:
$
\mathbb E \left[ \frac{1}{n}\sum_{i=1}^n 1_{\{U_i\,\mathrm{coll.}\}} \right] = \mathbb P(U_i\,\mathrm{coll.}).
$
Here, $\mathbb P(U_i\,\mathrm{coll.})$ is the probability that arbitrary fixed
user is collected, and the above equality holds by the users' symmetry.
Normalized throughput equals the expected number of collected users per
base station, per slot:
$
T(G)=\frac{1}{\tau\,m}\mathbb E \left[ \sum_{i=1}^n 1_{\{U_i\,\mathrm{coll.}\}} \right]
   = G\, \mathbb P(U_i\,\mathrm{coll.}).
$
Peak (normalized) throughput is the throughput maximized over all loads:
 $
T^\bullet(\delta):=\sup\{G \geq 0:\,T(G)\}.
 $
Given a $1-\epsilon$ coverage requirement, the maximal peak throughput
  $T^\star$ is the maximal value of $T^\bullet(\delta)$ over all
  $\delta$'s that obey the $1-\epsilon$ coverage ($1-\mathrm{exp}(-\delta) \geq 1-\epsilon$), i.e.,
  over all $\delta \geq \mathrm{ln}(1/\epsilon).$
We define the threshold load~$G^\star(\delta)$
 as the maximal normalized load $G$ for which $\mathbb P(U_i\,\mathrm{coll.})$  is still
 at the maximal possible value $1-\mathrm{exp}(-\delta)$ (i.e., PLR is still minimal possible, equal to~$\mathrm{exp}(-\delta)$), asymptotically:
 \begin{equation}
 \label{eqn-thr-load}
 G^\star(\delta) = \sup\{G \geq 0:\,\,\mathbb P(U_i\,\mathrm{coll.}) \rightarrow 1-e^{-\delta}\}.
 \end{equation}
If, for a certain decoding algorithm, it holds that $\mathbb P(U_i\,\mathrm{coll.})$ is less than $1-\mathrm{exp}(-\delta)$ for any (arbitrarily small) positive $G$,
we define $G^\star(\delta)=0.$


%
%
%


\vspace{-6mm}

\section{Decoding algorithms}
\label{sec-Algs}
We now present four decoding algorithms: 1) non-cooperative decoding; 2) spatial cooperation; 3) temporal cooperation; and 4) spatio-temporal cooperation. With the first two
decodings, we assume that users transmit in one uniformly randomly chosen slot per frame, i.e.,
$\Lambda_1 \equiv 1$; with the latter two decodings, users transmit
according to a distribution~${\Lambda}$. Throughout, we assume: 1) perfect packet replica decoding whenever a base station observes a singleton;
 and 2) perfect interference cancellation (both across slots and across base stations), and perfect packet replica decoding
 whenever cancelling the interference reveals a singleton.
 %
%
%
%
%

\emph{Non-cooperative decoding} is decoupled across slots; at each slot~$t$, each station $B_l$ collects a
user~$U_i$ if and only if $U_i$ is the only active user among the adjacent users of~$B_l$.
An example is shown in Figure~\ref{figure-primer-1}, the four left figures, top right. We can see
 that non-cooperative decoding collects one user--adjacent to three base stations.

\emph{Spatial cooperation} exploits the SIC mechanism across neighboring base stations. Whenever a base station detects
a singleton and collects a user, say user $U_i$, it sends the $U_i$'s message to all the other base stations that cover $U_i$. This allows for eliminating the contribution of $U_i$ in every superposition signal that contains $U_i$ and can therefore generate
 new singletons and new decoded users through an iterative recovery procedure.
 We assume that, at the beginning of decoding, each base station knows for each of its adjacent users~$U_i$ its ID, as well as which other base stations cover~$U_i$. (See also Section~\ref{section-discussion}.) This information can be acquired beforehand, e.g., through an association procedure. Also, we assume that any two base stations that have a common user can communicate via a dedicated link.
 Hence, no global (system-wide) knowledge or communication is necessary; a base station needs only the information from the system elements (users and base stations) that are physically close. Further, inter-base station communications
 are assumed to be inexpensive system resources. We now present decoding with spatial cooperation.
 It is decoupled across slots, i.e., one decoding algorithm is run after each time slot~$t$.
We henceforth focus on a single, fixed slot~$t$.
Decoding is iterative, and base stations operate over decoding iterations~$s$ in synchrony.
 We set the maximal number of iterations to~$m$. Namely, it can be shown that the algorithm does not progress further after~$m$ iterations are performed, i.e.,
 iterations $s > m$ do not yield additional collected users. (See ahead paragraph with heading
 Graph representation of decoding for an explanation why this is the case.)
 %
 Each station $B_l$
 maintains over~$s$ a signal $z_l=z_l(s)$ that serves as a current superposition
 signal. One iteration of decoding at $B_l$ is given in Algorithm~1.
%
%
%
\vspace{-4mm}
{\allowdisplaybreaks{\begin{algorithm}
\label{alg-spatial-coop}
\caption{One iteration of decoding with spatial cooperation at station~$B_l$}
\begin{algorithmic}[1]
\STATE \emph{(Check signal)}: $B_l$ verifies whether $z_l$ corresponds to a singleton. If so, it executes
 the collect and transmit step; otherwise, the receive and update step is performed.
\STATE \emph{(Collect and transmit)}: Station $B_l$ collects a user $U^{(l)}$ and recovers its ID.
 Subsequently, it transmits the message $x^{(l)}$ ($U^{(l)}$'s information packet and ID) to all the $B_k$'s, $k \neq l$, that are adjacent to $U^{(l)}$.
 Then, station~$B_l$ leaves the algorithm.
%
%
%
%
%
\STATE \emph{(Receive and update)}: Station~$B_l$ collects all the messages $x^{(k)}$ that it received at~$t$ and
forms the list~$\mathcal{J}^{(l)}$ of all distinct messages among the received messages; $B_l$
  subtracts from $z_l$ the interference contributions from all the $x_j$'s,  $j \in \mathcal{J}^{(l)}$,
  which we symbolize as $z_l \leftarrow z_l - \sum_{j \in \mathcal{J}^{(l)}}x_j$.
  Set $s \leftarrow s+1$. If $s=m$, $B_l$ leaves the algorithm; if $s<m$, $B_l$
  goes to step~1.
%
\end{algorithmic}
\end{algorithm}}}

\textbf{Graph representation of decoding}.
 Decoding at slot~$t$ can be represented
via evolution of a bipartite graph $\mathcal G$ over iterations~$s$. At iteration~$s=0$,
the graph $\mathcal G$ is initialized to graph $\mathcal G_0$, defined as follows: $\mathcal G_0$'s set of variable nodes is the set of all \emph{active}
 users at slot~$t$; its set of check nodes is the set of all base stations; and
 the set of links is the set of all pairs $(B_l,U_i)$, such that
 $B_l$ and $U_i$ are adjacent--lie within distance~$r$ (and $U_i$ is active).
  At iteration~$s$, $\mathcal G$ changes as follows. Visit all
  check nodes (in parallel), and remove from $\mathcal G$
 all the check nodes with degree one. Also, remove
  all their incident edges, all their adjacent variable nodes, as well as
  the adjacent variable nodes's incident edges.
    See Figure~\ref{figure-primer-1}, left four figures: the top right figure shows an example of
   the initial graph $\mathcal G_0$, and top right and bottom show
   the evolution of $\mathcal G$ along iterations~$s$. It is easy to see that the algorithm terminates after at most $m$ iterations. Namely, at each iteration~$s$, either at least one base station node is removed, or the algorithm terminates at~$t$. Therefore, at most $m$ iterations can be performed.

\textbf{Degree distributions in $\mathcal G_0$}. For subsequent analysis of non-cooperative decoding and spatial cooperation,
it is useful to determine the users' degree distribution in $\mathcal G_0$.
 Denote by $D_i$ the $U_i$'s spatial degree, i.e., the number of its adjacent base stations in $\mathcal G_0$.
 Let $\Delta_d := \mathbb P \left( D_i=d\,|\,u_i \in \mathcal A^{\mathrm{o},r}\right)$.
  It is easy to show that:
 $
  \Delta_d = {m \choose d} (r^2\pi)^d (1-r^2\pi)^{m-d},$ $
  d=0,...,m.$
%
%
%
%
%
In the asymptotic setting (See Subsection~\ref{subsection-perf-metrics}),
when $m r^2\pi \rightarrow \delta$, $\delta>0$, we have that
the boundary placements' effect vanishes, and:
 $
\mathbb P \left( D_i=d \right) \rightarrow e^{-\delta} \frac{\delta^d}{d!},$$ d=0,1,...
 $
That is, the users' (spatial) degree distribution in $\mathcal G_0$ is asymptotically a Poisson distribution
with parameter~$\delta$. Similarly, it is easy to show that a base station $B_l$'s
degree distribution in $\mathcal G_0$ is asymptotically Poisson with parameter $\delta\,G$, i.e.,
the probability that $B_l$ is adjacent to $d$ users converges to:
$
e^{-\delta\,G}\frac{(\delta\,G)^d}{d!},$ $d=0,1,...$

%

\emph{Temporal cooperation} utilizes the temporal SIC mechanism but is decoupled across base stations.
 Decoding at each frame is performed at the end of the frame (after users finish their transmissions).
Each base station runs, independently from other base stations, the standard (temporal) SIC over its (local) slots; see Subsection~\ref{subsection-single-base-stations}. A user $U_i$ is then collected if and only if it is collected after the SIC decoding at (at least) one of its adjacent base stations. %
%
%
%

\emph{Spatio-temporal cooperation} utilizes SIC both locally, across individual base stations' slots, and also
across the neighboring base stations. Each base station $B_l$, over decoding iterations, interleaves  the following two steps:
1) standard SIC over its local slots until there are no more singleton slots (temporal cleaning), and it subsequently sends the decoded users' messages to the base stations that share these users; and 2) for each received user~$U_i$, it cleans the $U_i$'s contribution at each of the $U_i$'s activation slots (spatial cleaning). The iterative decoding algorithm is done after all transmissions within the frame are completed and is done as follows. The number of iterations equals~$\tau m$. (It can be shown that no progress
is made at iterations $s>\tau m$.)
%
Each base station~$B_l$ performs the
same iterations~$s$; they are synchronous over all stations, i.e., the stations work in parallel.
 Station $B_l$ updates over iterations~$s$ the signals $z_{l,t}(s)$, where
$z_{l,t}(s)$ is the current superposition signal at slot~$t$. Note that now
 each base station $B_l$ maintains over iterations a set of $\tau$ signals $z_{l,t}(s)$, $t=1,...,\tau$.
  We detail iteration~$s$ at station $B_l$ in Algorithm~2. In step~1 \emph{(Temporal SIC and Transmit)} of Algorithm~2,
  station $B_l$ performs the standard temporal SIC across its local time slots, as explained in Subsection~\ref{subsection-single-base-stations}. (The maximal number of temporal SIC iterations can be limited to $\tau$ without loss in performance.)
%
%
%
%
{\allowdisplaybreaks{\begin{algorithm}
\label{alg-spatio-temporal-coop}
\caption{One iteration of decoding with spatio-temporal cooperation at station~$B_l$}
\begin{algorithmic}[1]
\STATE {\emph{(Temporal SIC and Transmit)}}: Station $B_l$ performs SIC across its local time slots and forms the list $\mathcal U^{(l),\mathrm{out}}$ of collected users during current temporal SIC. For each $U^{(l)}$ in $\mathcal U^{(l),\mathrm{out}}$, $B_l$ broadcasts the information packet from $U^{(l)}$, the $U^{(l)}$'s ID, and the $U^{(l)}$'s activation slots list, to all the base stations adjacent to~$U^{(l)}$. Perform step~2.
\STATE {\emph{(Check termination)}}: If either all the slots at station~$B_l$ are resolved or $s=\tau m$, $B_l$ leaves the algorithm. Else, it performs step~3.
\STATE
{\emph{(Receive and Spatial ICs)}}:  Station~$B_l$ makes the set $\mathcal U^{(l),\mathrm{in}}$  of all distinct users that it received at step~1. If $\mathcal U^{(l),\mathrm{new}}:=\mathcal U^{(l),\mathrm{in}} \setminus
\mathcal U^{(l),\mathrm{out}} = \emptyset$ (empty set), set $s\leftarrow s+1$ and perform step~2. Else, for each $U^{(k)}$ in $\mathcal U^{(l),\mathrm{new}}$, $B_l$ subtracts the contribution of $U^{(k)}$ at all its local slots where $U^{(k)}$ was active,
which we symbolize as $z_{l,t}\leftarrow z_{l,t} - U^{(k)}$.
 Set $s \leftarrow s+1$ and go to step~1.
%
%
\end{algorithmic}
\end{algorithm}}}

%

\textbf{Graph representation of decoding}.
We represent spatio-temporal cooperative decoding via evolution of a bipartite
graph $\mathcal H$ over iterations~$s$. At $s=0$, $\mathcal H$ is initialized to $\mathcal H_0$,
 defined as follows: $\mathcal H_0$'s set of variable nodes is the set of \emph{all users}; the set of check nodes
is the set of all pairs $(B_l,t)$, $l=1,...,m$, $t=1,...,\tau$; and the set of edges
is the set of all pairs $(U_i,(B_l,t))$, such that
$U_i$ and $B_l$ are adjacent (within distance~$r$), and
$U_i$ transmits at slot~$t$. Graph $\mathcal H$ evolves over iterations according to Algorithm~2.
 See Figure~\ref{figure-primer-1}, the right four figures, for an example of graph $\mathcal H$'s
 evolution over iterations~$s$.

\textbf{Degree distributions in $\mathcal H_0$}.
For subsequent analysis of spatio-temporal cooperation, it is useful
to determine the users' (variable nodes') and check nodes' degree distributions. Denote by $Z_i$ the degree of $U_i$ (arbitrary variable node) in~$\mathcal H_0$,
and recall the $U_i$'s temporal degree $Q_i$, and the $U_i$'s spatial degree $D_i$. Since all placements are fixed during the frame, whenever active, $U_i$ is heard by the same set of base stations. Therefore, $Z_i=D_i Q_i$.
  We do not pursue here directly the degree distribution, i.e.,
  we do not evaluate $\mathbb P(Z_i=d)$, $d=0,1,...$; instead, we will need its
  polynomial representation
  $\mathbb E\left[ x^{Z_i} \right]=\sum_{d=0}^{\infty}\mathbb P(Z_i=d) x^d$, $x \in [0,1]$.
   Conditioning on $Q_i$ and exploiting independence of $Q_i$ and $D_i$ (which follows from the independence of a user's activation from users' and base stations' placements), we have $\mathbb E\left[ x^{Z_i} \right]= \sum_{s=1}^{s_{\mathrm{max}}}
   \Lambda_s \mathbb E\left[ x^{s D_i}\right]$. Using the latter and the polynomial representation of $D_i$, it can be derived
   (it can be shown that the effects of boundary placements vanish) that $\mathbb E\left[ x^{Z_i} \right]$
   is asymptotically~(see~\cite{ISITarxiv} for details):
$
\Gamma(x) := \sum_{s=1}^{s_{\mathrm{max}}} \Lambda_s   e^{ -\delta (1-x^s)},$ $\forall x \in [0,1].
$
This is the asymptotic \emph{node-oriented} users' degree distribution.
 We will also need the edge-oriented distribution $\gamma(x)=\Gamma^\prime(x)/\Gamma^\prime(1)$, e.g.,~\cite{RichardsonUrbanke}.
 A straightforward calculation shows that:
\begin{equation}
\label{eq-characteristic-function-2}
\gamma(x) := \sum_{s=1}^{s_{\mathrm{max}}} \frac{s\,\Lambda_s}{\lambda} x^{s-1} e^{ -\delta (1-x^s)},\:\:\forall x \in [0,1],
\end{equation}
where we recall that $\lambda = \mathbb E[Q_i]=\sum_{s=1}^{s_{\mathrm{max}}} s \Lambda_s$.
 It can be shown (see \cite{ISITarxiv}; see also, e.g.~\cite{liva}) that the (edge-oriented) degree distribution $\chi(x)$ for arbitrary fixed check node~$(B_l,t)$ is asymptotically:
 \vspace{-3mm}
\begin{equation}
\label{eq-characteristic-function-3}
\chi(x) := e^{-G \delta \lambda(1-x)},\:\:\forall x \in [0,1].
\end{equation}

%
%

\vspace{-8mm}
\section{Performance analysis}
\label{section-performance-analysis}
This Section states our results on the four decoding algorithms:
non-cooperative (Subsection~\ref{subsection-results-non-coop}), spatial cooperation (Subsection~\ref{subsection-spatial-coop}),
temporal cooperation (Subsection~\ref{subsection-temporal-results}), and spatio-temporal
cooperation (Subsection~\ref{subsection-spatio-temporal-results}).

\subsection{Non-cooperative decoding}
\label{subsection-results-non-coop}
%
%
%
%
%
%
We first introduce certain auxiliary variables that play an important role in determining the
performance of non-cooperative decoding. Let $q_1,...,q_k$ be the points
generated uniformly at random (mutually independently) in the unit-area ball $\mathbf{B}(0,1/\sqrt{\pi})$. Let $\alpha_k$ be the area of the union
  $\cup_{s=1}^k \mathbf{B}(q_s,1/\sqrt{\pi})$. Further, denote by~$\mu_k$ the probability distribution of $\alpha_k$.
  Clearly, $\alpha_1 $ equals one with probability one, and $\mu_1$ is the delta distribution centered at one.
  Also, it is easy to see that, for any $k$, $\alpha_k \leq 4$, with probability one. It is also clear
  that the means $\overline{\alpha}_k$ are increasing in $k$, and lie between $1$ and $4$. Quantities~$\overline{\alpha}_k$'s can be obtained
  using Monte Carlo simulations~\cite{MASSAP2}.
 %
In Theorem~\ref{theorem-non-coop}, we characterize the decoding probability $\mathbb P \left( U_i\mathrm{\;coll.}\right)$ for both finite and asymptotic regimes.
\begin{theorem}[Non-cooperative: Decoding probability]
\label{theorem-non-coop}
Consider non-cooperative decoding. Then:
\vspace{-6mm}
\begin{enumerate}[(a)]
\item For $0<r \leq 1/4$:
 $
P_{\mathrm{coll.}}^{\,\mathrm{o},r}(1-4r)^2$ $\leq \mathbb P \left( U_i\mathrm{\;coll.}\right)\leq $ $ P_{\mathrm{coll.}}^{\,\mathrm{o},r} (1-4r)^2+8r-16r^2 ,
 $
  where
 $
P_{\,\mathrm{coll.}}^{\mathrm{o},r} = \mathbb P \left(U_i\mathrm{\;coll.}\,|\,U_i\mathrm{\;act.},\,u_i \in \mathcal{A}^{\mathrm{o},r}\right),
$
 and equals:
\begin{equation}
\label{eqn-P-zeta-k}
P_{\,\mathrm{coll.}}^{\mathrm{o},r} = \sum_{k=1}^m (-1)^{k-1}\,\zeta_k\,\int_{a=1}^4 \left(1- \frac{r^2\pi a}{\tau} \right)^{n-1}\,d\mu_k(a),\;\;\;\zeta_k= \sum_{d=k}^m {d \choose k}\,\Delta_d.
\end{equation}
\item Asymptotically, we have:
\begin{equation}
\label{eqn-asymptotic-non-coop}
\mathbb P \left( U_i\mathrm{\;coll.} \right) \rightarrow
\sum_{k=1}^{\infty} (-1)^{k-1}\,\frac{\delta^k}{k!}\,\int_{a=1}^4 e^{-\delta\,G\,a}\,d\mu_k(a)
\geq (1-e^{-\delta})e^{-\delta\,G}.
\end{equation}
\end{enumerate}
\end{theorem}
Proof of Theorem~\ref{theorem-non-coop} is in the supplementary material. 
We first comment on the structure of the results. The
integrals $\int_{a=1}^4 \left(1-r^2\pi a /\tau\right)^{n-1}\,d\mu_k(a)$ in~\eqref{eqn-P-zeta-k} converge to the integrals
 $\int_{a=1}^4 e^{-\delta\,G\,a}\,d\mu_k(a)$ in~\eqref{eqn-asymptotic-non-coop}. Also,
 $\zeta_k \rightarrow \frac{\delta^k}{k\!}$, and hence, as $r \rightarrow 0$ in the asymptotic setting,
 one can obtain the limit in~\eqref{eqn-asymptotic-non-coop} from~\eqref{eqn-P-zeta-k}.
 Obtaining the exact result with the alternating sum in~\eqref{eqn-P-zeta-k} is non-trivial
 and is obtained here using the inclusion-exclusion principle (See the supplementary material.)
   Also, note that, at $G=n/(\tau m)=0$ (number of users $n$ grows to infinity slower than $\tau m$), $\mathbb P(U_i\,\mathrm{coll.})$ equals
  the maximal possible value~$1-\mathrm{exp}(-\delta)$ asymptotically.

In practice, for $m$ of order $50$ or larger, the difficult-to-compute
 formula~\eqref{eqn-asymptotic-non-coop} can be approximated via the following easy-to-compute formula (see also~\cite{MASSAP2}):
  $
 \sum_{k=1}^{k_{\mathrm{max}}} (-1)^{k-1} \frac{\delta^k}{k!} e^{-\overline{\alpha}_k \,\delta \,G},
  $
 where recall $\overline{\alpha}_k$ is the mean of the distribution $\mu_k$ which can be estimated through
 Monte carlo simulations. 
  We remark that the $\overline{\alpha}_k$'s need to be estimated only once.
 Once we obtain them, they can be used for any set of system parameters~$n,m,\tau,r$. The quantity $k_{\mathrm{max}}$
 should be large enough relative to $\delta$; e.g., $k_{\mathrm{max}} \geq 5 \delta$. We proceed by establishing the achievable maximal peak throughput, maximized over all $\delta$'s that ensure $(1-\epsilon)$-coverage.
\begin{corollary}[Non-cooperative: Peak throughput]
\label{corollary-non-coop-peak-throughput}
Assume that the system has the $1-\epsilon$ coverage. Then, the quantity
$T^\star \geq\frac{1}{e}\frac{1-\epsilon}{\mathrm{ln}(1/\epsilon)}$. Hence, as $m$ grows large, the
unnormalized throughput (number of collected users per slot across all base stations)
is at least $\frac{1-\epsilon}{\mathrm{ln}(1/\epsilon)}\times m$ larger
than the throughput of the corresponding single base station system.
\end{corollary}
\begin{IEEEproof}
Suppose that $\delta \geq \mathrm{ln}(1/\epsilon)$, i.e., the $\epsilon$-coverage holds.
From Theorem~\ref{theorem-non-coop}, we have that, asymptotically,
 $
T(G) \geq T^\prime(G):=G\,(1-e^{-\delta})e^{-\delta\,G}.
 $
 Maximizing $T^\prime(G)$ over $G\geq 0$, we obtain:
 $
T^\star(\delta)
 \geq T^{\prime \prime}(\delta):=\frac{1-e^{-\delta}}{\delta\,e}.
 $
 The latter quantity is a decreasing function of $\delta$, and hence its
 maximum is attained at the minimal $\delta=\mathrm{ln}(1/\epsilon)$; substituting the latter value of $\delta$ in $T^{\prime \prime}(\delta)$, the result follows.
\end{IEEEproof}
From Theorem~\ref{theorem-non-coop}, we can easily obtain that the threshold load $G^\star(\delta)$
 is zero with the non-cooperative decoding.
\begin{corollary}[Non-cooperative: Threshold load]
\label{corollary-non-coop-threshold-load}
The threshold load~$G^\star(\delta)=0$. The decoding probability decreases at $G=0$ from the value $1-\mathrm{exp}(-\delta)$
 with the negative slope equal in magnitude to $\delta\,\sum_{k=1}^{\infty}(-1)^{k-1}\,\overline{\alpha}_k\,\delta^k/k!$.
\end{corollary}
\begin{IEEEproof}
The result follows by differentiating (more precisely, by taking the right derivative of) the sum in~\eqref{eqn-asymptotic-non-coop} with respect to $G$, and setting $G=0.$
\end{IEEEproof}
\vspace{-4mm}
\subsection{Spatial cooperation}
\label{subsection-spatial-coop}
We now turn our attention to spatial cooperation. By construction of the
non-cooperative and spatial algorithms, it is clear that
the decoding probability of spatial cooperation is greater than or equal the decoding
probability of the non-cooperative decoding. Hence, the non-cooperative decoding
probability is a lower bound on the spatial algorithm's decoding probability.
In Lemma~\ref{lemma-spatial-coop}, we devise an upper bound
on the spatial algorithm's decoding probability. The bound may be loose for larger $G$'s, but it
allows for establishing the threshold load $G^\star(\delta)$ with spatial cooperation.
Proof of Lemma~\ref{lemma-spatial-coop} is in the supplementary material.

\begin{lemma}[Spatial cooperation: Decoding probability upper bound]
\label{lemma-spatial-coop}
Consider decoding with spatial cooperation. Then, $\mathbb P (U_i\,\mathrm{coll.})$ is asymptotically upper bounded by:\footnote{
Here, the precise meaning of the wording asymptotically upper bounded is that $\limsup_{n \rightarrow \infty} \mathbb P (U_i\,\mathrm{coll.})
\leq 1-e^{-\delta}-(1-e^{-\delta/4})e^{-2\delta}(1-e^{-G\delta/4}).$ To keep the notation simple,
we will use this wording repeatedly throughout the paper.}
\begin{equation}
\label{eqn-spatial-lemma}
1-e^{-\delta}-(1-e^{-\delta/4})e^{-2\delta}(1-e^{-G\delta/4}).
\end{equation}
\end{lemma}
The upper bound in~\eqref{eqn-spatial-lemma} matches the actual spatial cooperation's performance at $G=n/(\tau\,m)=0$. (This corresponds to
the asymptotic setting when the number of users $n$ grows to infinity slower than $\tau m$.)
 Namely, note that, at $G = 0$, the quantity in~\eqref{eqn-spatial-lemma} equals
$1-\mathrm{exp}(-\delta)$. On the other hand, we have already shown that with the non-cooperative
decoding $\mathbb P(U_i\,\mathrm{coll.})$ is $1-\mathrm{exp}(-\delta)$ at $G = 0$.
   Hence, as $\mathbb P(U_i\,\mathrm{coll.})$ with spatial cooperation is
 larger than or equal to that of non-cooperative decoding,
 we conclude that, with spatial cooperation, $\mathbb P(U_i\,\mathrm{coll.})$ indeed equals
  $1-\mathrm{exp}(-\delta)$ at $G = 0$ and matches~\eqref{eqn-spatial-lemma}.
  However, from~\eqref{eqn-spatial-lemma}, we can see that, at arbitrarily small $G>0$,
  \eqref{eqn-spatial-lemma} is strictly smaller than $1-\mathrm{exp}(-\delta)$,
  and so is $\mathbb P(U_i\,\mathrm{coll.})$. This means that the threshold
  $G^\star(\delta)=0$. This conclusion is formalized in the following Corollary.
\begin{corollary}[Spatial cooperation: Threshold load]
\label{corollary-spatial-threshold}
The threshold $G^\star(\delta)=0$. The decoding probability decreases at $G=0$ from the value $1-\mathrm{exp}(-\delta)$
 with the negative slope, which is in magnitude at least equal to $\frac{1}{4}\,\delta\,\mathrm{exp}(-2\delta)(1-\mathrm{exp}(-\delta/4))$.
\end{corollary}
\begin{IEEEproof}
The proof follows by differentiating (more precisely, by taking the right derivative of) the quantity in~\eqref{eqn-spatial-lemma} with respect to
$G$, at $G=0$.
\end{IEEEproof}
We can see that, with spatial cooperation, although the performance is improved with respect to the non-cooperative case and an iterative decoding is employed,
we still have the zero threshold. This occurs due to the localized, geometric structure of~$\mathcal G_0$,
and the emergence of certain stopping sets (see, e.g.,~\cite{RichardsonUrbanke}) with a non-vanishing probability. (See the proof
of Lemma~\ref{lemma-spatial-coop} in the supplementary material.)
 \vspace{-4mm}

\subsection{Temporal cooperation}
\label{subsection-temporal-results}
We now consider temporal cooperation with temporal degree distribution~$\Lambda$. Recall from Subsection~\ref{subsection-single-base-stations} ${\rho}(H)$--the asymptotic decoding probability at load $H$ for
the single base station system with temporal SIC and the same temporal degree distribution~$\Lambda$.
%
%
%
%
\begin{theorem}[Temporal cooperation: Decoding probability lower bound]
\label{theorem-temporal-coop-decod-prob}
Consider temporal cooperation where users transmit according to
the temporal degree distribution~$\Lambda$. Further, assume the asymptotic setting in Subsection~\ref{subsection-perf-metrics}.~Then, decoding probability $\mathbb P (U_i\,\mathrm{coll.})$ is asymptotically lower
 bounded by $
(1-e^{-\delta}) \,{\rho}\left(H=(1+\epsilon)4 \delta G\right)$,
where $\epsilon>0$ is arbitrarily small.
\end{theorem}
Proof of Theorem~\ref{theorem-temporal-coop-decod-prob} is similar
to the proof of Theorem~1 in~\cite{ISITarxiv} and is in the supplementary material.
 Note the very interesting structure of the bound and the similarity with the lower bound in~\eqref{eqn-asymptotic-non-coop}.
The difference is that the standard slotted Aloha term $\mathrm{exp}(-H)$ at $H=\delta G$ is replaced
with the slotted Aloha with temporal SIC term ${\rho}(H)$ at $H = (1+\epsilon)(4 \delta G)$.

The next Corollary establishes existence of a non-zero threshold load $G^\star(\delta)$, and
it provides a lower bound on the threshold. The threshold lower bound is expressed
explicitly in terms of the single-base station threshold load $H^\star$ for the same distribution~$\Lambda$ and the users' average
spatial degree~$\delta$.
\begin{corollary} [Temporal cooperation: Threshold]
\label{corollary-temp-coop-threshold}
The threshold $G^\star(\delta)\geq \frac{1}{4}\frac{H^\star}{\delta}$.
Hence, the decoding probability stays at the maximal possible value $1-\mathrm{exp}(-\delta)$
at least in the range $G \in [0,\frac{1}{4}\frac{H^\star}{\delta}].$
\end{corollary}
\begin{IEEEproof}
Fix $\epsilon>0$. We know that, for the single base station system with temporal SIC,
it holds that $\rho(H)=1$ if $H\leq H^\star$. Hence, from Theorem~\ref{theorem-temporal-coop-decod-prob},
we have that $\mathbb P(U_i\,\mathrm{coll.}) \rightarrow 1-\mathrm{exp}(-\delta)$
 if $(4\,\delta\,G)(1+\epsilon) \leq H^\star$, i.e., if $G \leq \frac{H^\star}{4\,\delta(1+\epsilon)}$.
 By the definition of $G^\star(\delta)$ in~\eqref{eqn-thr-load}, it follows that
 $G^\star(\delta) \geq \frac{H^\star}{4\,\delta(1+\epsilon)}.$ Letting $\epsilon \rightarrow 0$,
 the desired result follows.
\end{IEEEproof}
Finally, the next Corollary establishes the achievable maximal peak throughput $T^\star$; the result is similar in spirit to Corollary~\ref{corollary-non-coop-peak-throughput}.
\begin{corollary}[Temporal cooperation: Peak throughput]
\label{corollary-temp-coop-peak-throughput}
Assume that the system has the $1-\epsilon$ coverage. Then, the quantity
$T^\star \geq \frac{H^\star}{4} \frac{1-\epsilon}{\mathrm{ln}(1/\epsilon)}$. Hence, as $m$ grows large, the
unnormalized throughput (number of collected users per slot across all base stations)
is at least $\frac{1}{4} \frac{1-\epsilon}{\mathrm{ln}(1/\epsilon)}\times m$ larger
than the throughput of the corresponding single base station system.
\end{corollary}
\begin{IEEEproof} Assume that $\delta \geq \mathrm{ln}(1/\epsilon)$, i.e.,
the~$1-\epsilon$ coverage holds.
Using the formula $T(G)=G\,\mathbb P(U_i\,\mathrm{coll.})$,
and the fact that, at $G=\frac{H^\star}{4\,\delta}$ we have that
$P(U_i\,\mathrm{coll.})$ is $1-e^{-\delta}$ asymptotically, we conclude that,
asymptotically, the peak throughput:
$
T^\bullet(\delta) \geq \frac{H^\star\,(1-e^{-\delta})}{4\,\delta}.
$
 We now maximize the latter function over~$\delta\geq \mathrm{ln}(1/\epsilon)$.
  We calculate the derivative of $\psi(\delta):=(1-\mathrm{exp}(-\delta))/\delta$,
  which  equals $\psi^\prime(\delta)=\frac{(1+\delta)\mathrm{exp}(-\delta)-1}{\delta^2}$.
  We show that $\psi^\prime(\delta) \leq 0$, for all $\delta \geq 0$.
   Indeed, the derivative of $(1+\delta)\mathrm{exp}(-\delta)$
   equals $-\delta \mathrm{exp}(-\delta) \leq 0$, $\forall \delta \geq 0$.
   Hence, $(1+\delta)\mathrm{exp}(-\delta) \leq (1+0)\mathrm{exp}(-0)=1$,
   which implies that $\psi^\prime(\delta) \leq 0$, $\forall \delta \geq 0$.
   Hence, $\phi(\delta)$ is non-increasing over $\delta \geq 0$.
   Hence, its maximum over $\delta \geq \mathrm{ln}(1/\epsilon)$
    is at $\delta = \mathrm{ln}(1/\epsilon)$.
    Finally, evaluating $\frac{H^\star\,(1-e^{-\delta})}{4\,\delta}$
     at $\delta= \mathrm{ln}(1/\epsilon)$ gives the~desired~result.
\end{IEEEproof}
%
%
%

%
%
%
\vspace{-6mm}
\subsection{Spatio-temporal cooperation}
\label{subsection-spatio-temporal-results}
We now study spatio-temporal cooperation. By the algorithm's construction, it is clear that the decoding probability with spatio-temporal cooperation is larger than or equal to decoding probability with temporal cooperation. Hence, all the results in
Subsection~\ref{subsection-temporal-results} continue to hold with spatio-temporal cooperation, as well. Next,
we give a heuristic for evaluation of the decoding probability.

\textbf{A heuristic for evaluating decoding probability}.
Exact evaluation of decoding probability~(PLR) with spatio-temporal cooperation is
a very challenging problem. However, we
 are able to calculate here the asymptotic degree distributions of
graph~$\mathcal H_0$ in closed form (see~\eqref{eq-characteristic-function-2}--\eqref{eq-characteristic-function-3}).
 This allows us to devise a heuristic based on
  and-or-tree iterations, e.g.,~\cite{liva}.
   %
    With spatial cooperation, we have observed numerically
    that and-or-tree iterations may yield over-optimistic estimates of
    the throughput and PLR. A major reason for this
    is the emergence of short cycles (and certain local stopping sets)
    with spatial decoding
    graph~$\mathcal G_0$. However, with spatio-temporal
    cooperation, the effect of these local stopping sets is
    reduced, causing that and-or-tree iterations
    give better performance predictions.
    See the supplementary material for an intuitive explanation of the latter effect.
    Given graph~$\mathcal H_0$,
    derivation of the and-or-tree equations is completely analogous
    to that in Section~{IV} of~\cite{liva},
    where the degree distributions $\Lambda(x)$, $\lambda(x)$, and $\rho(x)$
     in~\cite{liva} are now replaced with $\Gamma(x), \gamma(x)$, and $\chi(x)$, respectively.
    %
%
%
%
%
%
%
  Therefore, we estimate $\mathbb P (U_i\,\mathrm{coll}.)$ and $T(G)$ as
 \begin{equation}
 \label{eqn-P-dec-spatio-temporal}
 \mathbb P (U_i\,\mathrm{coll}.) \approx 1-\Gamma(p_{S}),\:\:T(G) \approx G\,(1-\Gamma(p_{S})),
 \end{equation}
 where $p_{S}$ is the output of the and-or-tree evolution, initialized by $p_0=q_0=1$, and iterations:
 $
q_s = \gamma(p_{s-1}),$ $p_s = 1-\chi(1-q_s),$ $s=1,...,S.$
We set the maximal number of iterations  $S=\tau\,m$.

\textbf{Threshold estimate}. We denote by $G^\bullet(\delta,{\Lambda})$ the and-or-tree
estimate of the spatio-temporal threshold load $G^\star(\delta,{\Lambda})$
  Following, e.g.,~\cite{RichardsonUrbanke}, $G^\bullet(\delta,{\Lambda})$ is obtained as the largest load $G$ for which:
$
f(G,{\Lambda};\,q)-q<0,$ $\forall q \in (0,1],$
where $f(G,{\Lambda};\,q) :=$$ \gamma \left( 1-e^{-G\delta\lambda\,q} \right).$
 (Recall that $\lambda = \sum_{s=1}^{s_{\mathrm{max}}} s \Lambda_s$ is the users' average temporal degree.)
A simple upper bound on $G^\bullet(\delta,{\Lambda})$ is obtained from the stability condition, e.g.,~\cite{liva}.
The condition says that, at $G=G^\bullet(\delta,{\Lambda})$, there must hold that
$\frac{d f(G,{\Lambda};\,q)}{d\,q}\,|_{q=0} \leq 1.$ After differentiation and simple algebraic
manipulations, the stability condition yields:
 $
G^\bullet(\delta,{\Lambda}) \leq \frac{e^{\delta}}{\delta}\,\frac{1}{2\,\Lambda_2}\,\frac{1}{1+\frac{\delta \Lambda_1}{2\Lambda_2}} $ $\leq \frac{e^{\delta}}{\delta}\,\frac{1}{2 \Lambda_2}.
 $
Note that the term $\frac{1}{2\Lambda_2}$ is an upper bound on the single-base station system threshold~$H^\star$
obtained from the stability condition~\cite{liva}.

\textbf{Optimization of the temporal degree distribution~$\Lambda$}. Given $m$ and $r$ (equivalently, given $\delta=m r^2\pi$),
we seek ${\Lambda}=(\Lambda_1,...,\Lambda_{s_{\mathrm{max}}})^\top$, that maximizes $\phi({\Lambda}):=G^\bullet(\delta,{\Lambda})$ over
all probability distributions $\Lambda$ defined on the $s_{\mathrm{max}}$-dimensional alphabet. This is a challenging optimization problem.
However, in practice, $s_{\mathrm{max}}$ is typically assumed small, e.g., $s_{\mathrm{max}}=8$,~\cite{liva},
and it is feasible to numerically perform optimization.
 We employ the following algorithm to maximize~$\phi({\Lambda})$. For a fixed ${\Lambda}$, we numerically
estimate~$\phi({\Lambda})$ as follows. We discretize the interval $q \in (0,1]$ with
$J$ equidistant points, $q_j=j/J$, $j=1,...,J$, and we estimate $\phi(\Lambda)$ as:
\begin{equation}
\label{eqn-function-f-approx}
\max\{ G\geq 0:\,\max_{j=1,...,J}\left(f(G,{\Lambda};\,q_j)-q_j\right)<0 \}.
\end{equation}
The function~$\max_{q \in (0,1]}\left(f(G,{\Lambda};\,q)-q\right)$ is
 non-decreasing in~$G$;
 hence, we calculate~\eqref{eqn-function-f-approx} via
the bisection method.
 As, given ${\Lambda}$, we can (approximately) evaluate~$\phi({\Lambda})$, we can apply
a gradient-free numerical optimization procedure to find an optimal~${\Lambda}$. We use a variation of
the iterative, random optimization method in~\cite{RandomOptimization}.


\vspace{-8mm}

\section{Numerical studies}
\label{section-simulations}
We now perform numerical optimization for the users' temporal degree
distribution with spatio-temporal cooperation, and we demonstrate by simulation
 the validity of our optimization method. We also show by simulation
 that spatio-temporal cooperation yields significant
 gains in terms of peak throughput and PLR when compared with the remaining three schemes.

%
%

\textbf{Simulation setup}. We set the number of base stations $m=40$,
and the number of slots $\tau=40$ (unless stated otherwise). We simulate decoding probability $\mathbb P(U_i\mathrm{\,coll.})$
 versus $G=n /(\tau m)$ by varying $n$. We perform Monte Carlo simulations. For each value of $n$,
 we generate $\mathrm{MC}=30$ instances of the network ($30$ placements of users and base stations) with all the methods except spatio-temporal
 cooperation, where we run $\mathrm{MC}=300$ instances due to lower achieved PLRs.
 For each placement, we run the decoding algorithms. For each $n$ (each $G$), we estimate
$\mathbb P(U_i\mathrm{\,coll.})$ as $\frac{1}{n}\frac{1}{\mathrm{MC}}\sum_{s=1}^{\mathrm{MC}}N_s$,
where $N_s$ is the number of collected users for the $s$-th random placement.
With temporal and spatio-temporal cooperation, simulations include the following distributions:
1) $\Lambda_2 \equiv 1$, proposed in~\cite{SlottedALOHAwithIC}; 2)
the single-base station optimized distribution in~\cite{liva}:
$\Lambda_2=0.5$, $\Lambda_3=0.28$, $\Lambda_8=1-\Lambda_2-\Lambda_3$; and 3)
optimized distributions as explained in Section~\ref{section-performance-analysis}. With non-cooperative decoding and spatial cooperation, we simulate the distribution~$\Lambda_1\equiv 1$.
 When comparing different decodings in terms of PLR, we set the target PLR values from the
 following set: $\{0.01; 0.02; 0.1\}$. These values
 are practical and correspond to operation of LTE-A~\cite{LTEadvanced,3GPPMeeting}.
 Namely, reference~\cite{LTEadvanced}
 indicates a target PLR of $0.01$ for control channel, and $0.1$ for data channel, while~\cite{3GPPMeeting} indicates a target PLR of $0.02$.

\textbf{Spatio-temporal cooperation}. We now focus on spatio-temporal cooperation and
the effect of the users' temporal degree distribution~$\Lambda$. Due to practical considerations,
we set the maximal degree $s_{\mathrm{max}}=8$ as in~\cite{liva}. For the values~$\delta \in \{0.1, 0.3, 0.5, 1, 2, 3, 5, 7\}$,
 we optimize ${\Lambda}$ as explained in Section~\ref{section-performance-analysis}.
   Table~1 shows the obtained optimized distributions ${\Lambda^{\bullet}}$ for $\delta \in \{0.1, 0.3, 0.5, 1, 2,3, 5, 7\}$, rounded at two decimal places.
  %
  %
  {\begin{tiny}{
\begin{table*}\centering
\begin{tabular}{@{}lllllllll@{}}
\toprule
$\delta=$ & $\:\:\:\:\:\:0.1$ & $\:\:\:\:\:\:0.3$ & $\:\:\:\:\:\:0.5$ & $\:\:\:\:\:\:\:\:1$ & $\:\:\:\:\:\:\:\:2$ & $\:\:\:\:\:3$
& $\:\:\:\:\:\:\:\:5$ & $\:\:\:\:\:\:\:\:7$
\\
$\Lambda^{\bullet}=$ & $\left[ \begin{array}{cccccccc}
0 \\ 0.54 \\ 0.26 \\ 0.01 \\ 0 \\ 0.01 \\ 0 \\0.18 \end{array} \right] $
 & $\left[ \begin{array}{cccccccc} 0\\
   0.62\\
   0.20\\
                   0\\
                   0\\
                   0\\
   0.09\\
   0.09\end{array} \right]$
 & $\left[ \begin{array}{cccccccc}0\\
   0.68\\
   0.17\\
                   0\\
   0\\
   0\\
   0\\
   0.15\end{array} \right]$
 & $\left[ \begin{array}{cccccccc} 0\\0.91\\0\\0\\0\\0\\0\\0.09 \end{array} \right]$
 & $\left[ \begin{array}{cccccccc} 0\\1\\0\\0\\0\\0\\0\\0 \end{array} \right]$
 & $\left[ \begin{array}{cccccccc} 0\\1\\0\\0\\0\\0\\0\\0 \end{array} \right]$
 & $\left[ \begin{array}{cccccccc} 0.01\\0.99\\0\\0\\0\\0\\0\\0\end{array} \right]$
 & $\left[ \begin{array}{cccccccc} 0.10\\0.90\\0\\0\\0\\0\\0\\0\end{array} \right]$    \\
\bottomrule
\end{tabular}
\vspace{1mm}
\caption{Optimized $\Lambda^\bullet$ for different values of users' average
spatial degree~$\delta$.}
\vspace{-12mm}
\end{table*}}
\end{tiny}}
We can see that, for a very small~$\delta=0.1$, $\Lambda^\bullet$ is very close
to the single-base station optimal distribution in~\cite{liva},
equal to $(0.5, 0.28,0,0,0,0,0,0.22)^\top$. This is intuitive, as at small $\delta$'s,
 base stations' coverage regions do not overlap with high probability, and
  hence each base station works as an isolated single base station system.
  As we increase $\delta$, $\Lambda^{\bullet}$ becomes very close to the constant-degree-two
  distribution in~\cite{SlottedALOHAwithIC}. Moreover, for $\delta \geq 2$, the entries $\Lambda^{\bullet}_s$, $s\geq 3$, are all zero.
   Hence, we fine-tune the optimization by restricting to two-dimensional distributions~$(\Lambda_1,1-\Lambda_1)^\top$,
   for $\delta \in \{1,2,...,7\}$, and performing a one-dimensional grid search over $\Lambda_1 \in [0,1]$.
      The fine-tuning agrees with the results in Table~1
    for $\delta <7$; for $\delta=7$, the fine-tuning gave the constant-degree-two distribution.

\begin{figure}[thpb]
      \centering
      \vspace{-7mm}
       \includegraphics[height=2.1 in,width=3.1 in]{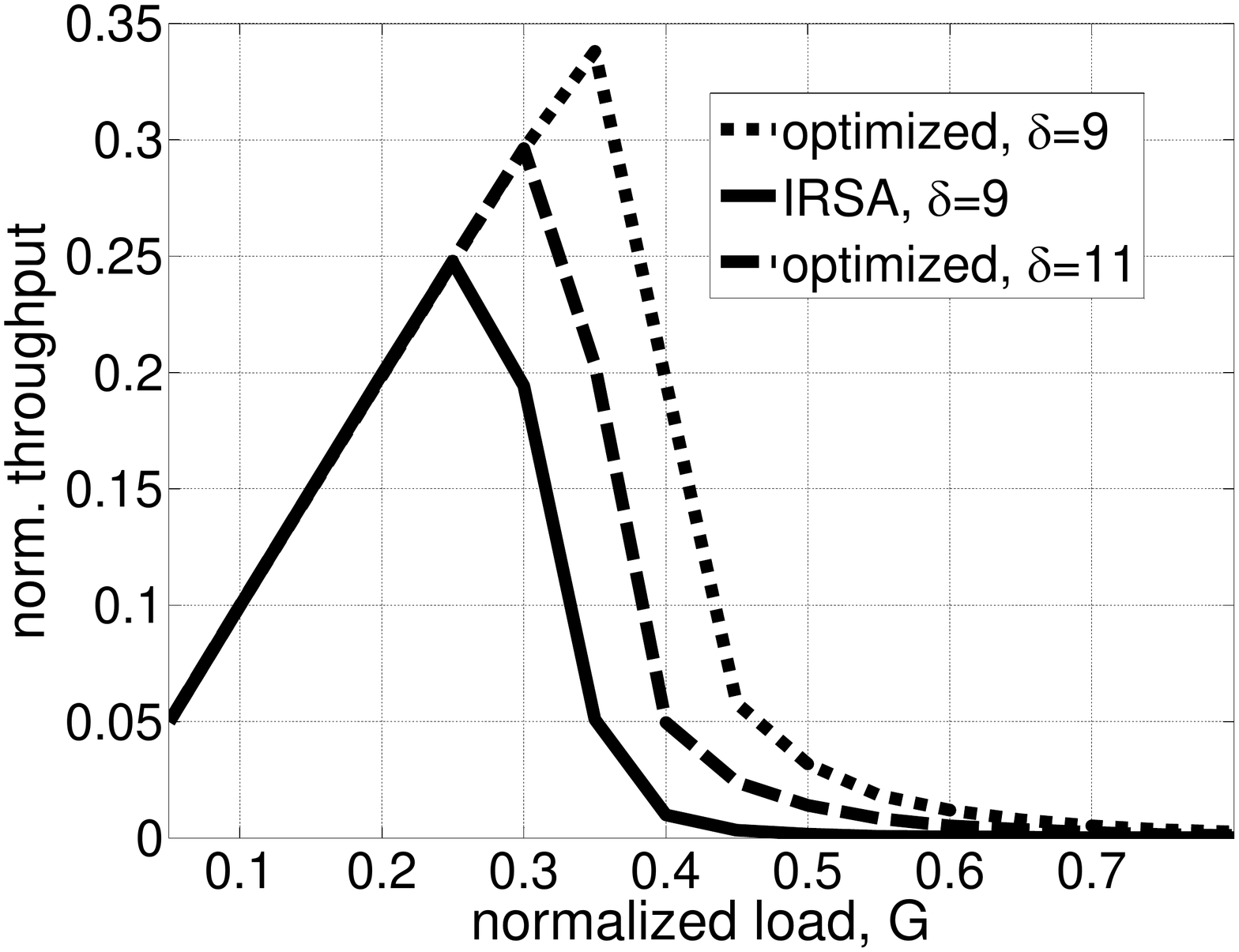}
       \includegraphics[height=2.1 in,width=3.1 in]{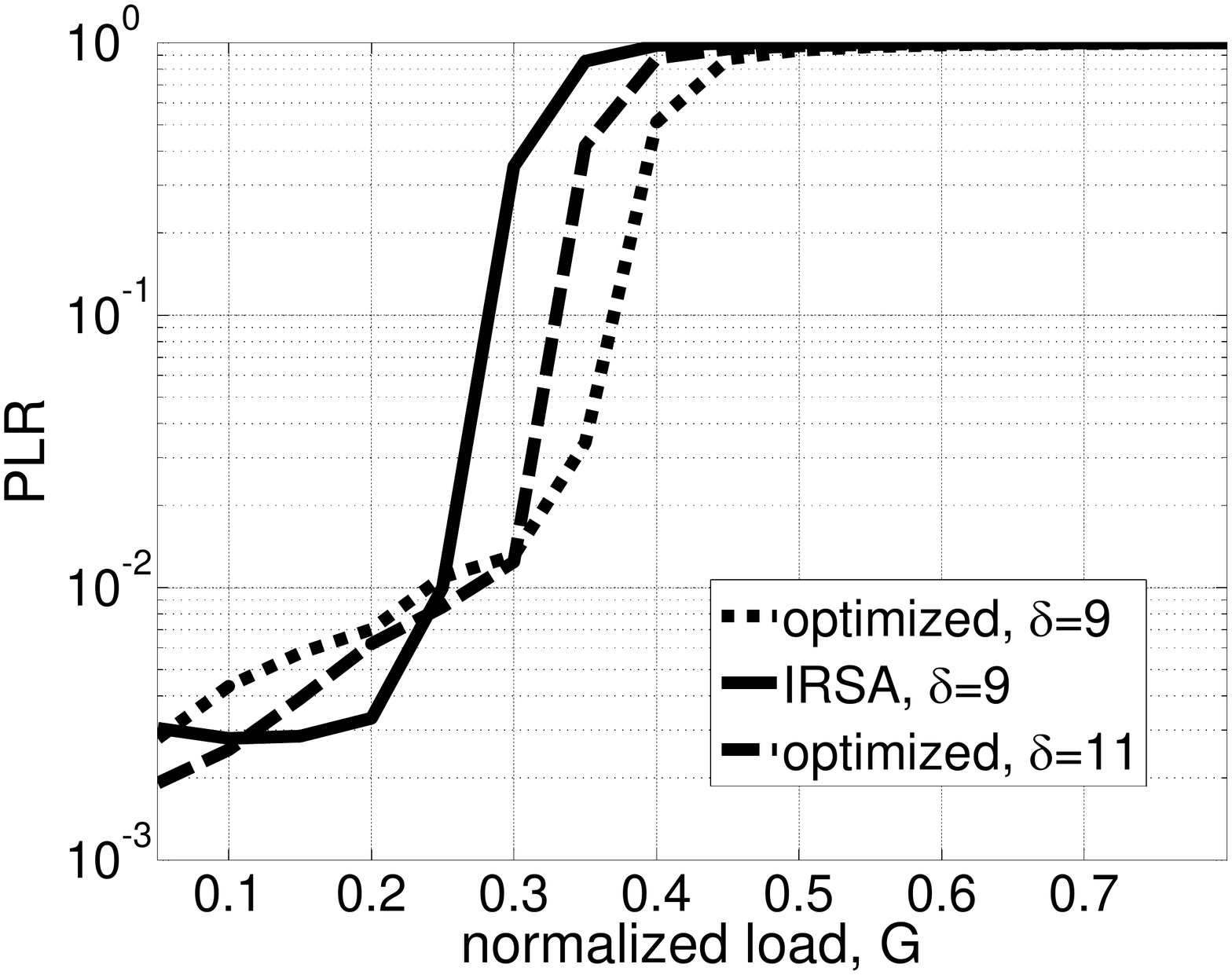}
       \vspace{-4mm}
       \caption{Left: Simulated normalized throughput $T(G)$ versus normalized load $G=n/(\tau m)$ for
       spatio-temporal cooperation. Right: Simulated PLR versus $G$ for
       spatio-temporal cooperation. The figures show the
       performance of our optimized $\Lambda^{\bullet}$ with $\delta=9$ (dotted line) and
       $\delta=11$ (dashed line),
       and the distribution in~\cite{liva} (IRSA) for $\delta=9$ (solid line).}
       \label{figure-simul-spatio-temporal}
       \vspace{-7mm}
\end{figure}
\begin{figure}[thpb]
      \centering
      \vspace{-7mm}
       \includegraphics[height=2.1 in,width=3.1 in]{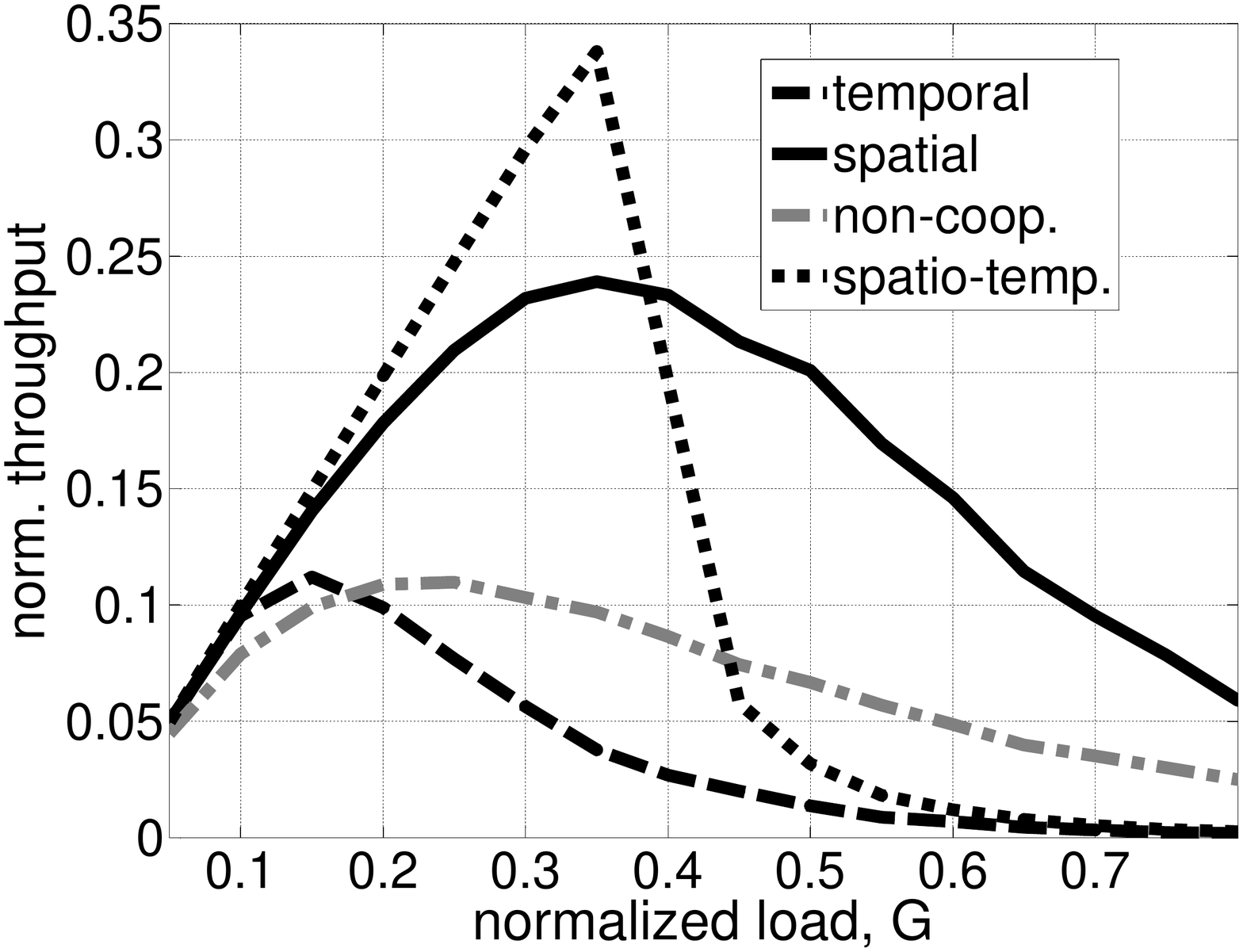}
       \includegraphics[height=2.1 in,width=3.1 in]{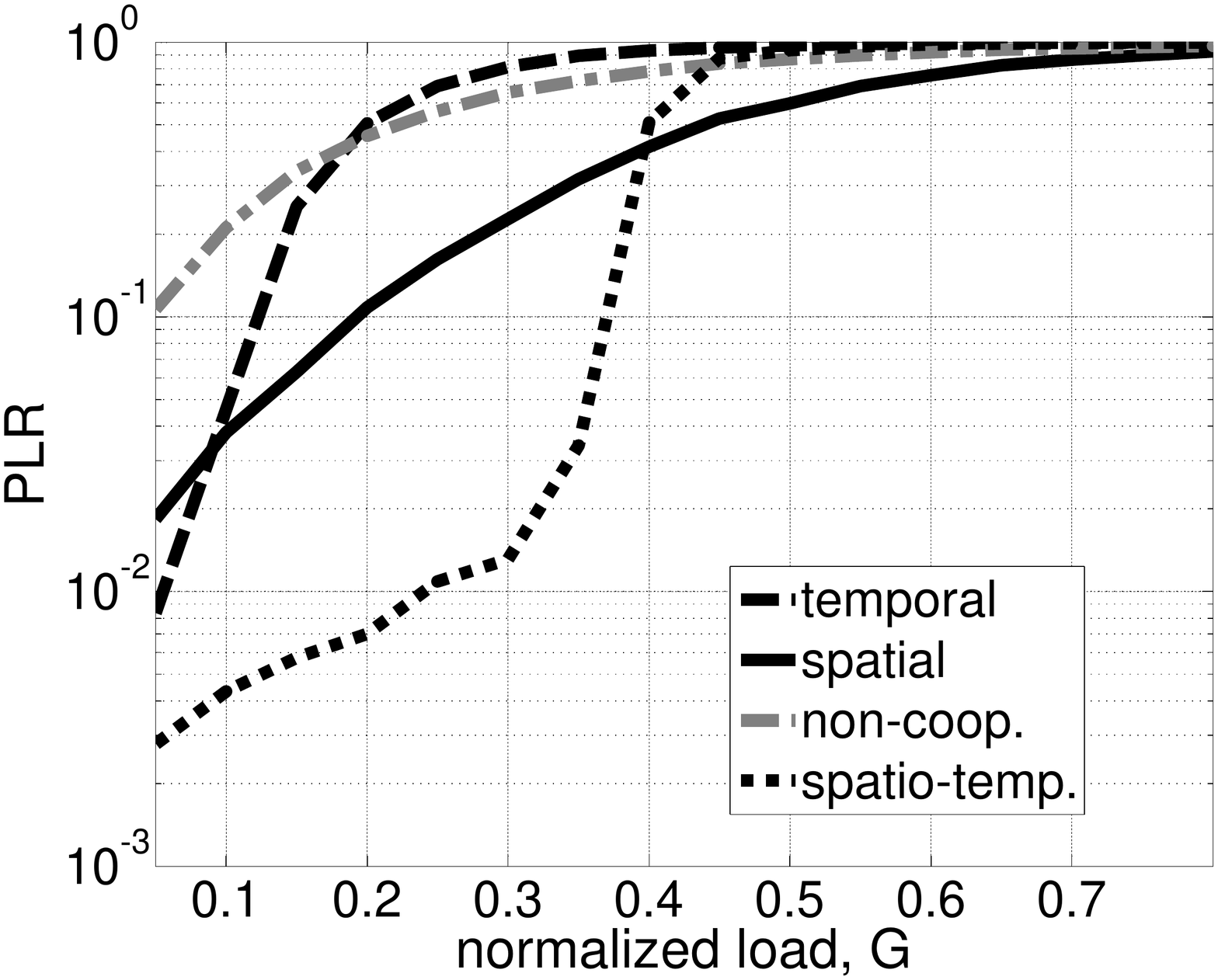}
       \vspace{-4mm}
       \caption{
        Performance of
        non-cooperative decoding (grey line), spatial cooperation (solid), temporal cooperation (dashed), and
        spatio-temporal cooperation (dotted) with the optimized $\Lambda^{\bullet}$ ($\Lambda^{\bullet}_2=1$,
        $\Lambda^{\bullet}_s=0$, $s \neq 2$), for $\delta=9$. Left: normalized throughput $T(G)$ versus normalized load $G$;
        Right: PLR versus normalized load $G$.}
       \label{figure-optimal-lambda-g-star}
       \vspace{-7mm}
\end{figure}
Figure~\ref{figure-simul-spatio-temporal} (left) plots normalized throughput $T(G)$ versus normalized load $G$ for
 $\delta=9$ (asymptotic minimal PLR$\approx 0.00012$) for
our (multi-base station optimized distribution) $\Lambda^{\bullet}$ and the single-base station optimized distribution in~\cite{liva}~(IRSA). For this value of $\delta$, the optimized distribution equals the constant-degree-two distribution.
We can see that~$\Lambda^{\bullet}$ indeed performs
better than~\cite{liva} in terms of the peak throughput ($0.34$ with $\Lambda^\bullet$
 versus $0.24$ with \cite{liva}), thus corroborating
our optimization method. In Figure~\ref{figure-simul-spatio-temporal} (right), we
 compare the two methods in terms of PLR (for both methods, $\delta=9$).
  For the target PLR of $0.1$,~$\Lambda^{\bullet}$ achieves it at the maximal load
   $G=0.37$, while~\cite{liva} achieves the target PLR at $G=0.28$.
   Similarly, for the target PLR of $0.02$, the maximal load with $\Lambda^{\bullet}$ is $0.32$,
   while with~\cite{liva} it is $0.26$. For the target PLR$=0.01$, the two methods perform almost the
   same,~\cite{liva} being slightly better (maximal load of $0.25$ with~\cite{liva} versus
   $0.24$ with $\Lambda^{\bullet}$.) This is a consequence of the non-asymptotic regime.
 At very small loads,
 both methods achieve asymptotically ($m \rightarrow \infty$) the same PLR--equal the
 minimal possible value $\mathrm{exp}(-\delta)\approx 0.00012$. Hence, asymptotically, as $G$ increases from zero,
 both methods start with PLR$\approx 0.00012$, maintain this value until the threshold load, and then start to increase PLR.
  (Note that our method has the larger asymptotic threshold load.)
  However, at a finite~$m$, the methods do not achieve asymptotic PLR.
  Also, at small loads $G \in [0.05,0.25]$, \cite{liva}
 achieves a better PLR. This means that~\cite{liva}
 approaches asymptotic performance faster (in~$m$) than our optimized method.
  This non-asymptotic effect reduces as $m$ becomes larger--the scenario highly relevant with M2M communications.
     For a given $m$ and a small target PLR, we can increase radius~$r$, i.e., increase~$\delta$ (with some
    additional resources spent) with our optimized distribution
   so that $\Lambda^\bullet$ achieves the target PLR at a larger maximal load than~\cite{liva}
   while still
   having a better throughput performance.
   Concretely, Figure~\ref{figure-simul-spatio-temporal} (right) additionally shows PLR for
   $\Lambda^\bullet$ and $\delta = 11$. We can see that, for the increased~$r$,
   $\Lambda^\bullet$ achieves the target PLR of $0.01$ at
   the maximal load $0.27$, while the corresponding
   maximal load with~\cite{liva} is $0.25$.
   Note from Figure~\ref{figure-simul-spatio-temporal} (left),
      that, at the same time, the peak
      throughput of our method with $\delta=11$ is larger than the
      peak throughput of~\cite{liva} with $\delta=9.$ Also, at load $G=0.27$ (operating point of $\Lambda^\bullet$ for the $0.01$ target PLR),
      the throughput with $\Lambda^\bullet$ is $0.27$, while with~\cite{liva} it is smaller and equals $0.22.$

\textbf{Comparison of the four decoding algorithms}. Figure~\ref{figure-optimal-lambda-g-star} (left) plots
 normalized throughput $T(G)$ versus normalized load $G$ for
non-cooperative decoding, spatial cooperation, temporal cooperation, and spatio-temporal cooperation,
for $\delta=9$. We can see that spatio-temporal cooperation achieves much higher peak normalized throughput ($\approx 0.34$)
than the remaining three schemes (spatial~$\approx 0.24$, temporal~$\approx 0.11$, and non-cooperative~$\approx 0.11$).
 Figure~\ref{figure-optimal-lambda-g-star} (right) compares the methods under the same parameters in terms of PLR.
 We can see that spatio-temporal cooperation performs significantly better than the remaining three schemes for each of the target PLRs. For example, for the target PLR$=0.02$, spatio-temporal cooperation achieves it at the
 maximal load $G=0.32$, temporal at $G=0.08$, spatial at $G=0.06$, while
 with the non-cooperative decoding the maximal load is below $G=0.05$.

\vspace{-5mm}
\section{Discussion}
\label{section-discussion}
In this Section, we include a discussion about the assumptions
that we make in the paper. We first explain how slot-synchronization and spatial SIC
 can be achieved in practice. Then, we discuss several aspects
 of the physical layer that are abstracted from our model. We also point to interesting
 future research directions.

\textbf{Slot-synchronization}.
We have assumed that users and base stations are synchronized
with respect to common slots. This can be, for example, achieved as follows.
   We can assume that all base stations periodically receive global positioning system--GPS markers of absolute time, and hence,
    they are all well-synchronized to absolute time.
    Prior to initiating a random access protocol, base stations agree on the frame length~$\tau$, time duration of each slot,
    and the instance of the absolute time when to initiate each frame.
    (This can be achieved, e.g., through the backhaul communication.) At the time instance
    of a frame start, all base stations broadcast to users the beacons that initiate the frames and contain the slot duration and frame length~$\tau$.

Propagation delays and the corresponding time offsets--assuming the above clock-synchronization of base
stations--will have a rather small effect in typical applications. For example, for a low-bit-rate M2M service in small-cell networks, if the worst-case difference in user-to-BS distances (among any pair of neighboring users of a base station) is 300 meters, the delay difference is on the order of 1 microsecond. This is typically less than the symbol period for a 100 kilobits-per-second service rate (where the bit period is 10 microseconds, while the symbol period might be longer if higher modulation constellations are used). (See also~\cite{NeighborDiscovery} for a similar discussion.)

The slot-synchronization assumption is also reasonable due to other evolving concepts that require tight neighboring base-station synchronization. For example, in LTE-A, neighboring base stations will require tight synchronization established via X2 interface. This is due to the requirements set by Coordinated Multi-Point (CoMP) functionality, where two or more neighboring base-stations collaborate in signal design in order to improve the received signal-to-interference-plus-noise-ratio (SINR) of cell edge users~\cite{CoMP}.
 For example, the differential delay among the packets addressed to different base stations is expected to be of order $1-5$ microseconds~\cite{LTEATiming}.

It is certainly relevant to also consider scenarios without slot synchronization.
   References~\cite{Kissling1,Kissling2,Gaudenzi2} develop asynchronous Aloha protocols with SIC.
   An interesting research direction is to develop such protocols for multi-base station systems as well.

\textbf{Interference cancellation}. We have assumed
perfect spatial and temporal interference cancellation.
   We first discuss spatial interference cancellation. We explain how spatial SIC can be achieved on an example where, at slot~$t$,
  $U_1$ is adjacent to $B_1$ and $B_2$,
  $B_2$ observes a singleton (and hence collects~$U_1$ and passes
  the $U_1$'s packet to $B_1$),
  while $B_1$ observes a collision. In order for $B_1$
  to subtract the $U_1$'s interference contribution,
  it needs estimates of the amplitude, phase offset, and frequency offset at slot~$t$~\cite{SlottedALOHAwithIC}. 
     With temporal SIC on satellite fixed channels~\cite{SlottedALOHAwithIC}, phase offset is estimated
   via preamble, directly at the collided slot, while
   amplitude and frequency offsets are copied from the clean burst~\cite{SlottedALOHAwithIC}.
     Here, the situation with phase and frequency can be considered analogous, but the amplitude
    needs to be estimated in a different way. This is because
    the amplitudes of the $U_1$'s signals at $B_1$ and $B_2$ are certainly different due to
    different distances from $U_1$ to $B_1$ and $B_2$, respectively (and perfect power control is not present).
       We take advantage of the fact that, in practice,
      the amplitude information can be available as a side information.
      For example, in LTE, users can measure the received signal power (averaged across the frequency bandwidth in use) of surrounding base stations using RSRP (Received Signal Reference Power) measurements of resource elements that carry cell-specific reference signals~\cite{RSP}. Hence, it
      is reasonable to assume that each user $U_i$
       has available channel gains $\gamma_{il}$ to all its
       adjacent base stations~$B_l$.
        Then, spatial SIC can be implemented as follows.
        Each $U_i$'s transmission packet contains the channel gains $\gamma_{il}$'s
        of its neighboring stations. In our example,
        after $B_2$ collects $U_1$, it reads off
        the channel gain $\gamma_{12}$ and passes this information
        to $B_2$, which is then able to subtract the $U_1$'s
        interference contribution.

        In situations when RSRP may not be available, amplitude, phase and frequency offsets
can be in principle estimated via the preamble. (Note that now the preamble serves to estimate
the latter three parameters, not only the phase offset as in~\cite{SlottedALOHAwithIC}.)
 Assume that each $B_l$ knows the preambles of all of its adjacent users.
  The received preamble at $B_l$ is then:
   $
  y_l = \sum_{j \in O_l} \zeta_j\, \gamma_{jl} \, e^{{\imath} (\phi_{jl} + \omega_{jl} \mathcal{T})} \mathcal{X}_j^{\mathrm{pre}} + \nu_l.
  $
  Here, ${\imath}$ is the imaginary unit, $O_l$ is the set of users $U_j$ adjacent to $B_l$ (both active and inactive); $\gamma_{jl}$, $\phi_{jl}$, and $\omega_{jl}$ are the amplitude,
 phase offset, and frequency offset, and
 $\mathcal{T}$ is the time instance of the current slot. (For notational simplicity,
  we dropped the dependence on slot~$t$.) Further,
   $\zeta_j$ is the Bernoulli random variable
   which indicates whether $U_j$ is active at the slot;
   $\mathcal{X}_j^{\mathrm{pre}}$ is the vector of preamble symbols of $U_j$; and $\nu_l$ is additive noise.
   Denote by $\eta_j := \zeta_j\, \gamma_{jl} \, e^{\imath(\phi_{jl} + \omega_{jl} \mathcal{T})}$,
   and by $\mathcal{X}^{(l)}$ the matrix whose columns
   are the vectors $\mathcal{X}_j^{\mathrm{pre}}$, $j \in O_l$. Then,
    the preamble equation is rewritten as:
    $
    y_l = \mathcal{X}^{(l)} \, \eta^{(l)} + \nu_l,
    $
    where $\eta^{(l)}$ is the vector that collects
    the~$\eta_j$'s, $j \in O_l$.
    Station $B_l$ can now obtain $\eta^{(l)}$ 
    via a standard linear estimation procedure.
     In our example, once $B_1$
     estimates $\eta^{(1)}$ (and hence, it has available $\eta_1$ that corresponds to $U_1$)
     and obtains the $U_1$'s information packet $\mathcal{X}_1$ from $B_1$, it can eliminate
     the interference contribution from $U_1$
     by subtracting~$\eta_1 \, \mathcal{X}_1$ from its signal.
 Vector~$\eta^{(l)}$ is usually sparse (due to sparse users' activation
at each slot), so it can be estimated 
via compressed-sensing type methods.

We now consider temporal interference cancellation. For satellite
fixed channels, references~\cite{SlottedALOHAwithIC,liva} demonstrate
a good performance of temporal SIC based on
copying the amplitude and frequency offset from the clean burst and
determining the phase offset
directly at the colliding burst. This technique is
based on the assumption that the amplitude and frequency (approximately) do not change
over different slots within a frame. This assumption may not be
adequate for terrestrial channels. In such scenarios,
we can estimate the channel amplitude, phase offset, and frequency offset
via the linear estimation method explained above.

Finally, it is an interesting future research direction to incorporate
the residual interference into the system model, as, e.g., done in a different context in~\cite{NonIdealSIC}. To our best knowledge, such analysis has not been
done yet even with SIC-Aloha single-base station systems.

\textbf{Base stations' knowledge of users neighborhoods}. With
spatial and spatio-temporal decodings, we have assumed that, at the beginning of decoding, each base station knows for each of its adjacent users~$U_i$ its ID, as well as which other base stations cover~$U_i$. This information can be acquired beforehand, e.g., through an association procedure. We also explain possible alternatives. First, note that,
 the only reason for requiring the above knowledge is that, when a station $B_l$ collects a user $U_i$,
it needs to send the $U_i$'s packet to other base stations adjacent to $U_i$. This can be achieved as follows. Recall that it is
reasonable to assume that users posses RSRP signals~\cite{RSP}, and hence they know the list of their adjacent base stations (the once whose RSRP exceeds a threshold.) Now, we let each user's transmission packet contain the list of all its adjacent base stations. Then, whenever a station~$B_l$ collects a user $U_i$, $B_l$ reads off the list of the $U_i$'s adjacent base stations, and hence the decoding algorithms can proceed as before. Another alternative is that, assuming users' placements are fixed within several frames, base stations in the initial frames work in a non-cooperative mode, employing non-cooperative or temporal decoding. Recall that these schemes do not require the users' IDs. Hence, through the initial frames, base stations can learn the IDs of (most of) their users, and subsequently switch to a cooperative mode (spatial or spatio-temporal).

\textbf{Physical layer model}. In this paper,
  we have assumed a MAC layer model which abstracts
several aspects of the physical layer. This is a common approach in random access and specially slotted Aloha with SIC, e.g.,~\cite{SlottedALOHAwithIC,liva,LivaInfoTheory,LivaNovo,FramelessALOHA,CedomirAlohaRateless}.
 It is worth noting that this paper (with our prior papers~\cite{MASSAP2,ISITarxiv,EWpaper}) and~\cite{Gaudenzi3} (where the latter does not provide analytical studies) are pioneering works on slotted Aloha with SIC for \emph{multi-base station systems}.
 As such, our paper naturally focuses on the MAC model.
  Analytical and detailed numerical studies of the physical layer are interesting future
research directions. Here, we provide a simulation example
under a physical layer model that accounts for several effects
 including path loss, fading/shadowing, and power unbalance. We demonstrate that
 the fundamental results and conclusions
 that we establish under the simpler model in Section~\ref{section-model-prel}
 are well-confirmed under this more detailed model also. Namely, we show: 1)
 linear increase in throughput with~$m$; 2) our optimized temporal degree
 distribution with spatio-temporal cooperation performs better than IRSA in~\cite{liva};
  and 3) threshold behavior continues to exist, i.e., PLR stays at a small
  value in a range of loads~$(0,G^\star]$.


We describe the model and extend spatio-temporal decoding to the novel setup. (Extension of the remaining three decodings is analogous.) The time slots and frame models, as well as the transmission protocol, remain the same as in Section~\ref{section-model-prel}, but the models of the received signal as well as
 of the base stations' decoding power are changed. A station~$B_l$ receives
 at slot~$t$ a superposition of the signals from \emph{all active users} at~$t$.
 The power of the contribution of $U_j$ is:
 $
 \mathcal{P}_{jl}(t) = \frac{\mathcal{P}_j \, g_{jl}(t)}{r_{jl}^{\alpha}}.
 $
  Here, $\mathcal{P}_j$ is the $U_j$'s transmit power; $\alpha$ is the path loss exponent; and $r_{jl}$ is the distance between $U_j$ and $B_l$. Further, $g_{jl}(t)$ is the fading/shadowing attenuation,
  modeled the same as in~\cite{HaenggiShadowing}, i.e., $g_{jl}(t)$ is a product of
  two independent random variables: an exponential
 variable with mean~$1$ (Rayleigh fading), and a log-normal
 variable whose natural logarithm is a standard normal variable (log-normal shadowing).
  The $g_{jl}(t)$'s are assumed independent, identically
 distributed across all triples~$j,l,t$. Users adopt power control with respect to
 their strongest base station; that is, $\mathcal{P}_{j} = (r_j^{\mathrm{min}})^{\alpha}$,
 where $r_j^{\mathrm{min}}$ is the distance to the station closest to~$U_j$.\footnote{The distance to the closest station can be estimated, e.g., via RSRP signals~\cite{RSP}; see the above paragraph with Heading Spatial interference cancellation.} Note that
 we still have power unbalance due to the fact that the
 $U_j$'s distance from different stations~$B_l$ is different, as well as due to fading.


For the purpose of defining the decoding algorithm, we introduce the base stations' coverage radius~$r$. Fix an arbitrary pair~$B_l,U_j$. Radius~$r$ is defined as the
largest distance~$r^\prime$ between $B_l$ and $U_j$ at
which the expected signal-to-noise ratio (SNR) (conditioned on $r_{jl}=r^\prime$) exceeds threshold~$\theta>0$:
\begin{equation}
\label{eqn-r-review-formula}
r =
\sup\left\{ r^\prime \geq 0:\,\mathbb E \left[\frac{\mathcal{P}_{jl}(t)/r_{jl}^{\alpha}}{\mathcal N}\,|\,r_{jl}=r^\prime\right] \geq \theta \right\},
\end{equation}
where $\mathcal N$ is the noise power, and the expectation is over the users' and base stations' placements and fading.
 In words, $r$ is the maximal distance at which, if $U_j$ is the only active user,
$B_l$ can still decode it (on average).
The parameter~$r$ depends on
$\mathcal N$, $\alpha$, $\theta$, and $m$, and can be estimated through Monte Carlo simulations.
We remark that this model still has certain simplifications. For example, in a realistic scenario, threshold $\theta$ is dependent on the speed of a
mobile user. The adopted model is more suitable for either stationary or low-mobility users.


The decoding bipartite graph~$\mathcal{H}_0$ is defined as before: there is a link between
check node $(B_l,t)$ and user~$U_j$ (variable node) if
and only if $U_j$ is active at~$t$ and the distance between
$U_j$ and $B_l$ is less than~$r$.\footnote{Clearly, this does not
mean that the $U_j$'s signal does not affect the signal of $(B_l,t)$ if their distance is beyond~$r$.
It only means that, if a check node $(B_l,t)$ (station $B_l$) collects a user~$U_j$, then
the $U_j$'s contribution is subtracted from the check nodes which
are adjacent to~$U_j$ in $\mathcal H$ (and is not subtracted from the remaining check nodes.)}
 %
%
The decoding algorithm is as follows. At each decoding iteration~$s$, each check node $(B_l,t)$ collects a user if its
  current SINR exceeds the threshold:
  \begin{equation}
 \label{eqn-review-threshold}
  \frac{\mathcal{P}_{il}(t)} {\mathcal{N} + \sum_{j \neq i, \,j \in O_l(t,s)} \mathcal{P}_{jl}(t)} \geq \theta.
  \end{equation}
Here, $O_l(t,s)$ is the set of users which are active at slot~$t$, and whose interference contribution is not
removed from the signal at check node $(B_l,t)$ up to iteration~$s$; and $i$ indexes the user in set~$j \in O_l(t,s)$ with highest power~$\mathcal{P}_{jl}(t)$ (strongest un-decoded user
 at check node $(B_l,t)$ and iteration~$s$). If~\eqref{eqn-review-threshold} is satisfied, the contribution from~$U_i$ is subtracted from
all check nodes in the current graph $\mathcal H$ adjacent to~$U_i$. (We still assume perfect
interference cancellation.) 

\begin{figure}[thpb]
      \centering
      \vspace{-7mm}
       \includegraphics[height=2. in,width=3. in]{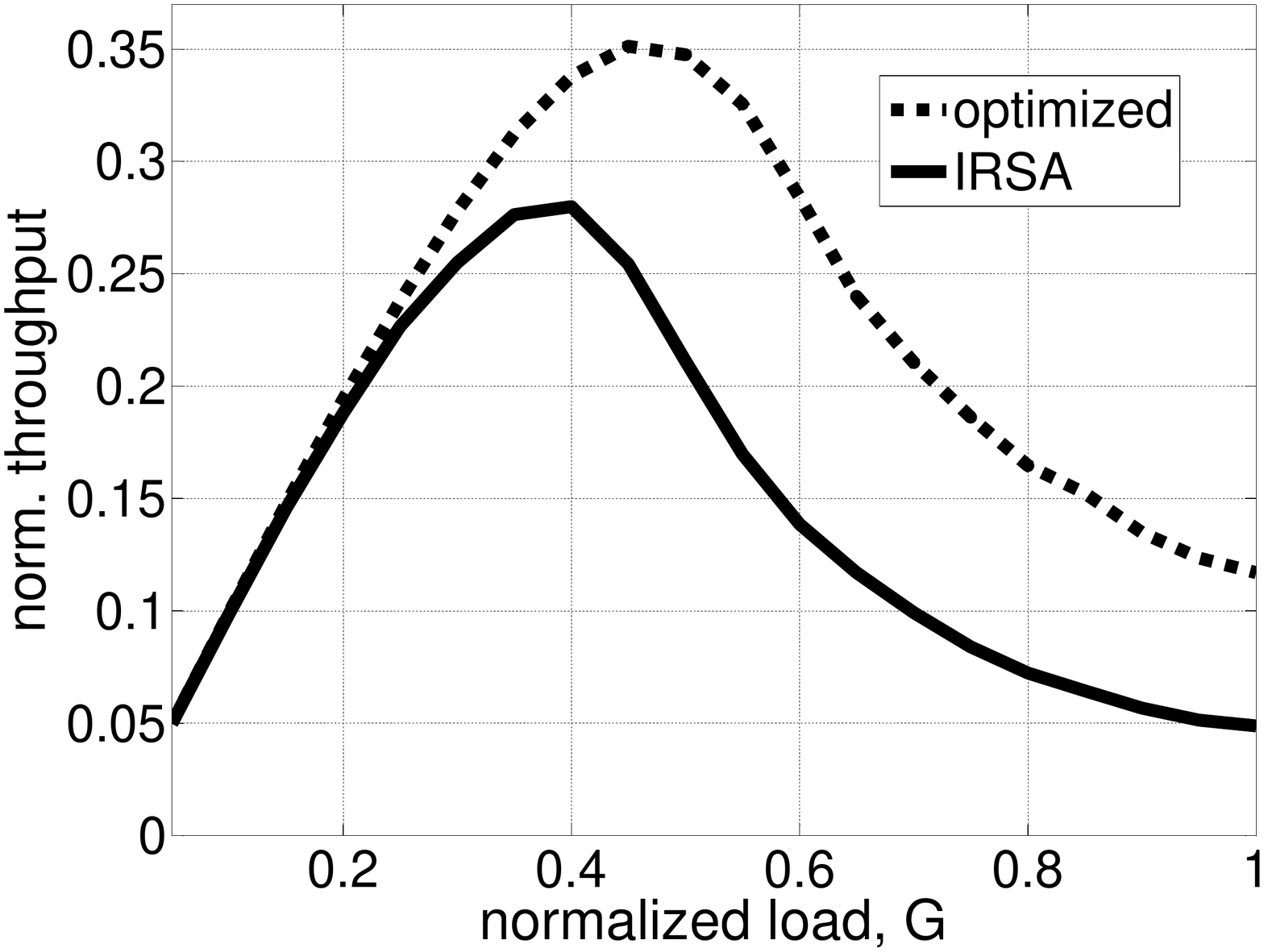}
       \includegraphics[height=2. in,width=3.2 in]{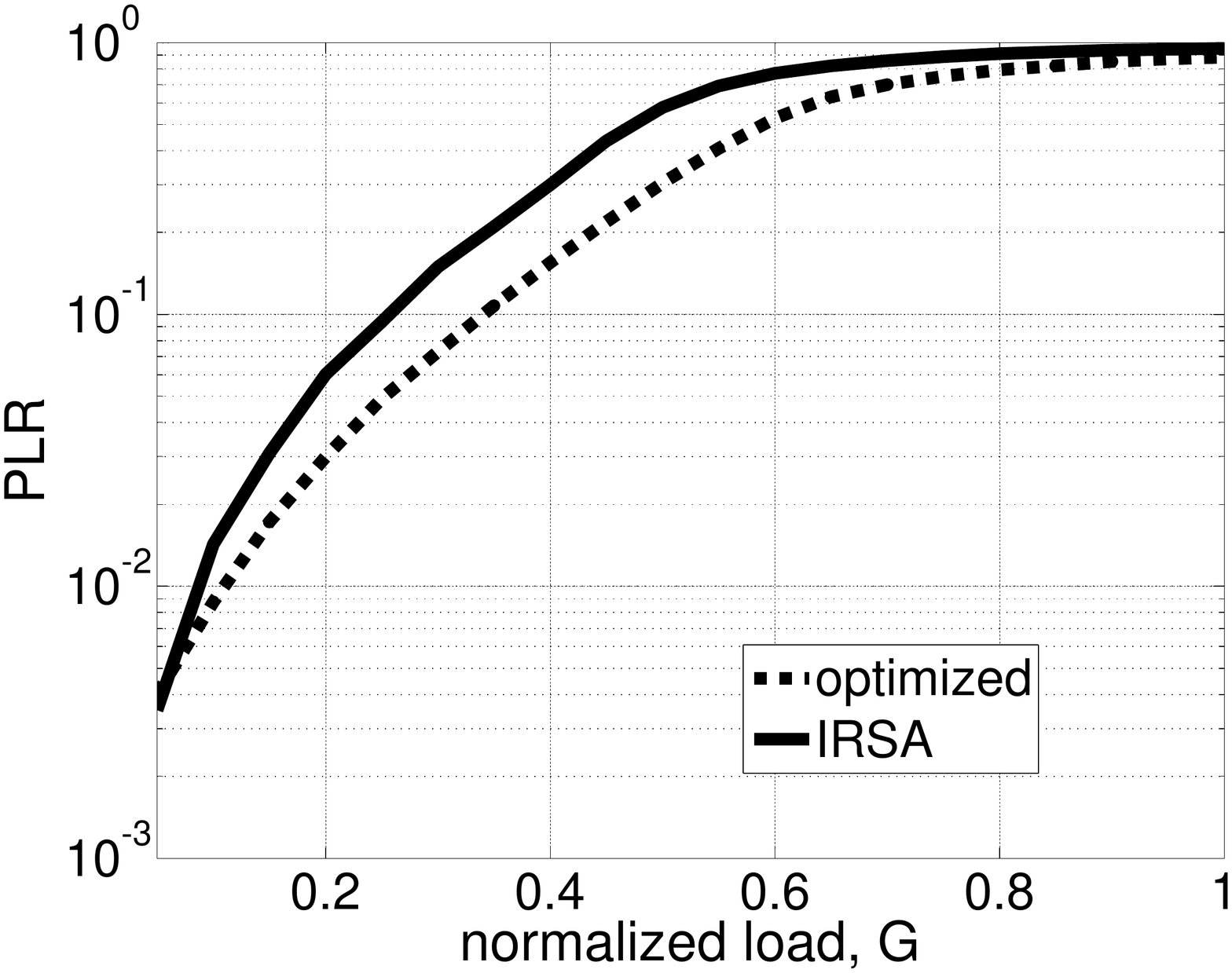}
       \vspace{-5mm}
       \caption{Comparison of the optimized degree distribution $\Lambda^\bullet$ and
       IRSA in~\cite{liva} for spatio-temporal cooperation on the physical layer model with noise power $\mathcal N=0.09$. Left: normalized throughput $T(G)$ versus
       normalized load~$G$; Right: PLR versus~$G$.}
       \label{figure-review-1}
       \vspace{-7mm}
\end{figure}

Simulation setup is as follows. There are $m=40$ base stations, $\tau=20$ slots per frame,
 path loss exponent $\alpha=2$,
and SINR threshold $\theta=1$. This threshold value
 corresponds approximately to the threshold decoding level for a robust (say binary phase shift keying--BPSK) modulation and a moderate (say half-rate) forward error correction--FEC option of the LTE physical layer (single-antenna) specifications.
 Noise power is $\mathcal N = 0.09$;
the corresponding
estimated radius $r = 0.39$ ($\delta=m r^2 \pi \approx 19.1$).
 Figure~\ref{figure-review-1} (left) plots the
normalized throughput versus normalized load~$G$ for our optimized
degree distribution $\Lambda^\bullet$ (equal the constant-degree-two distribution) and~\cite{liva}.
 We can see that $\Lambda^\bullet$
 achieves a higher peak throughput ($0.35$ with $\Lambda^\bullet$
 versus $0.28$ with~\cite{liva}). Figure~\ref{figure-review-1} (right)
  plots PLR versus $G$ for the two methods.
  We can see that $\Lambda^\bullet$ achieves a higher
  maximal load than~\cite{liva} for each target PLR.
  Specifically, the maximal loads for
  $\Lambda^\bullet$ and~\cite{liva} are, respectively:
  $0.11$ and $0.09$ (PLR$=0.01$);
  $0.16$ and $0.12$ (PLR$=0.02$); and
  $0.34$ and $0.26$ (PLR$=0.1$).
  We can see that the gain of our method with respect to~\cite{liva} is larger for larger target PLRs.

\begin{figure}[thpb]
      \centering
      \vspace{-0mm}
      \includegraphics[height=2. in,width=3.1 in]{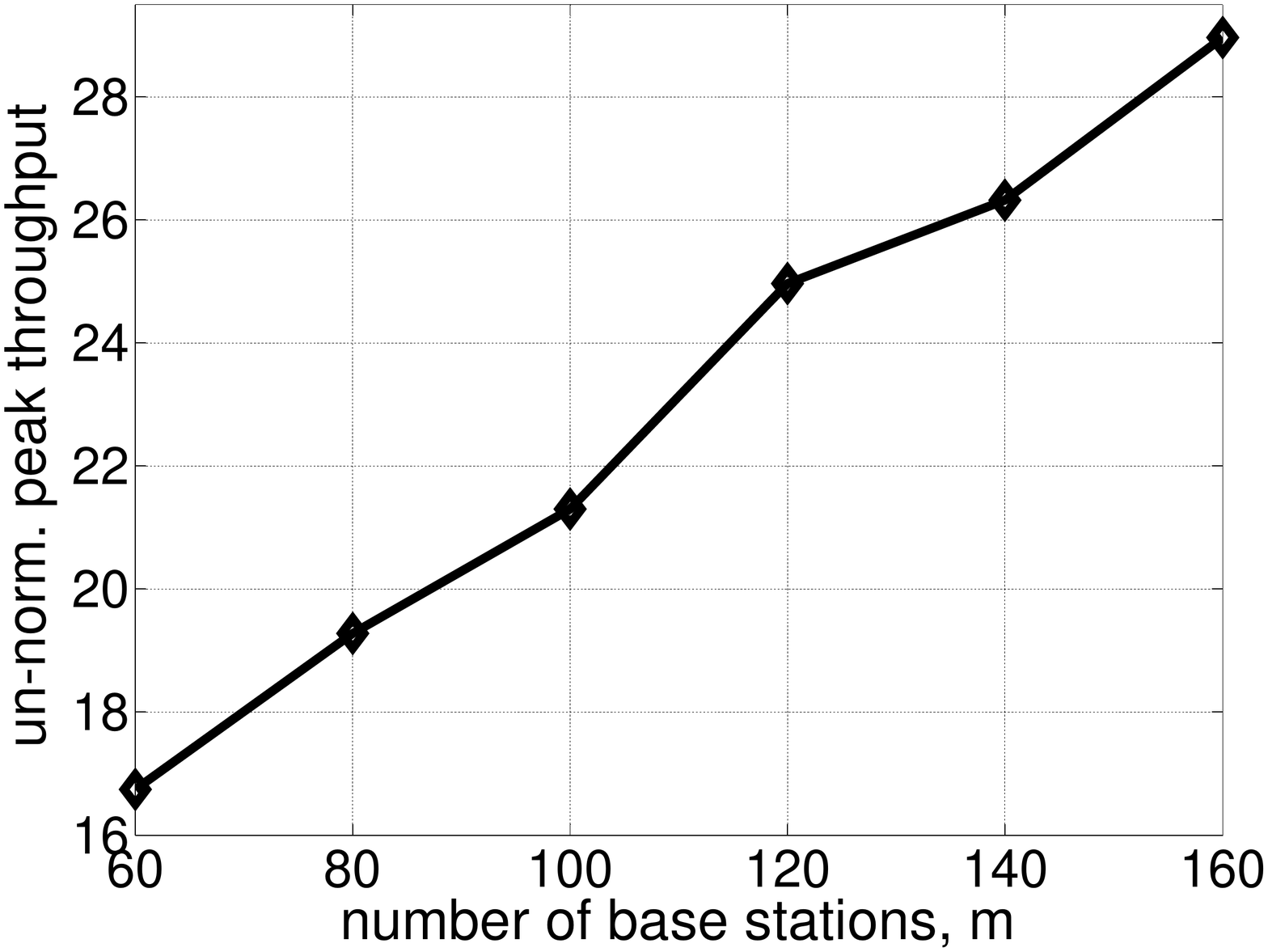}
       \includegraphics[height=2. in,width=3.2 in]{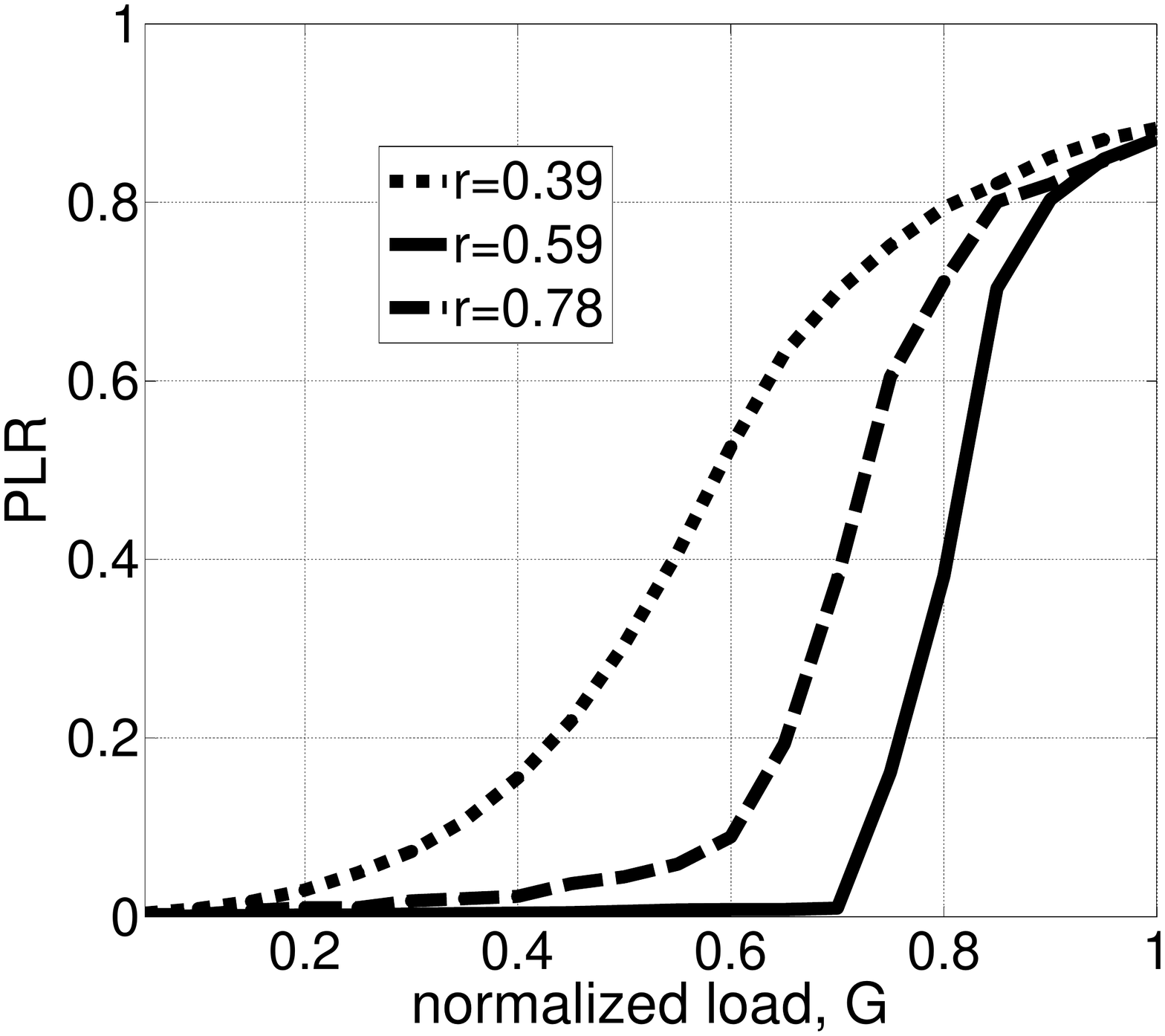}
       \vspace{-2mm}
       \caption{
       Spatio-temporal cooperation on the physical layer model with noise power $\mathcal N=0.09$.
       Left: Un-normalized peak throughput versus number of base stations~$m$
        for the optimized degree distribution $\Lambda^\bullet$. Right: PLR versus normalized
        load $G$ for $\Lambda^\bullet$ and different values of radius $r$;
        $0.39$ (dotted); $0.59$ (dashed); and $0.78$ (solid).
       }
       \label{figure-review-2}
       \vspace{-7mm}
\end{figure}

Figure~\ref{figure-review-2} (left) plots the aggregate peak throughput (expected number
of decoded users per slot, across all stations) versus~$G$ for $\mathcal N = 0.09$.
 We can see that it approximately increases linearly with~$m$, confirming our theory.
 Finally, we examine the effect of increasing base stations' cooperation (increasing
radius~$r$) while \emph{keeping} the same noise power~$\mathcal N = 0.09$; see Figure~\ref{figure-review-2} (right).
We consider $r = 0.39$ (obtained from~\eqref{eqn-r-review-formula}), $r = 0.59$, and $r=0.78$.
 We can see that, by increasing cooperation, the performance naturally improves,
 but also the threshold effect becomes more pronounced.

\vspace{-7mm}
\section{Conclusion}
\label{section-conclusion}
Recent works, e.g.,~\cite{SlottedALOHAwithIC,liva}, significantly improved the throughput of
standard slotted Aloha protocol by incorporating the successive interference cancellation (SIC)
mechanism into decoding process. In this paper, we extended~\cite{SlottedALOHAwithIC,liva} to the case
of multiple, cooperative base stations. We considered a geometric-proximity communication model
and proposed decoding algorithms that utilize either spatial or temporal cooperation, or both.
Spatial cooperation allows for interference cancellation across base stations, at a given slot,
while temporal cooperation allows for SIC across different slots. Specifically, we considered four decoding
algorithms: non-cooperative, spatial cooperation, temporal cooperation, and spatio-temporal cooperation,
and established several fundamental results on their performance.
 We showed that all algorithms have a linear increase of throughput (expected number of decoded users per slot, across all base stations) in the number of base stations, and we characterized the threshold load--the load below which the decoding probability equals the coverage probability of a fixed user. We found that temporal and spatio temporal cooperation exhibit a strictly positive threshold load, while
  non-cooperative decoding and spatial cooperation have zero threshold. Finally,
  with spatio-temporal cooperation, we optimized the users' temporal degree distribution. We showed
   that, when the system parameters are in a range of practical interest, the optimum is very different from the optimal transmission
   protocol when only one base station is present, and is close or equal to the constant-degree-two distribution.

\textbf{Acknowledgement}. We would like to thank anonymous reviewers for suggesting a significant addition to the paper, which improved it considerably.

\vspace{-7mm}
\bibliographystyle{IEEEtran}
\bibliography{IEEEabrv,bibliography}

\newpage
\section*{Supplementary material}
\subsection{Proof of Theorem~\ref{theorem-non-coop}}
 We first prove part~(a). Consider an arbitrary fixed user~$U_i$. Note that $U_i$ is active in exactly one of
the $\tau$ slots, equally likely across slots, and it can be decoded only if it is active.
Hence, using the total probability law,
 $\mathbb P(U_i\,\mathrm{coll.}) = \sum_{t=1}^{\tau}\mathbb P(U_i\,\mathrm{coll.}\,|\,U_i\,\mathrm{is\,active\,at\,t}) (1/\tau)
 = \mathbb P(U_i\,\mathrm{coll.}\,|\,U_i\,\mathrm{is\,active\,at\,1}) \sum_{t=1}^{\tau}(1/\tau)=
 \mathbb P(U_i\,\mathrm{coll.}\,|\,U_i\,\mathrm{is\,active\,at}\,1)$, where we used the symmetry across all slots.
 Hence, it suffices to consider slot~$t=1$, and find
 $\mathbb P(U_i\,\mathrm{coll.}\,$$|\,U_i\,\mathrm{is}$$\mathrm{\,act.\,at}\,1)$,
 which we will write simply as $\mathbb P(U_i\,\mathrm{coll.}\,|\,U_i\,\mathrm{is\,active})$.
 Let $U_i$ be placed at an arbitrary nominal placement~$q\in \mathcal A^{\mathrm{o},r}$.
  Denote by $\mathcal M(q)$ the subset of the indexes of
 the base stations that belong to $\mathbf{B}(q,r)$. Suppose that $u_i=q$ and $\mathcal M(q)= \mathcal I$, $\mathcal I \subset \{1,...,m\}$, $\mathcal I \neq \emptyset$. Then, $U_i$ is collected if at least one base station in $\mathcal I$ has no other active users besides $U_i$. Let $\mathcal B_l$ denote the (random) Euclidean ball of radius $r$ centered at the position of the base station $l$, i.e., $\mathcal B_l = \mathbf B(b_l,r)$, for $l=1,...,m$. For a base station $l$ that has no active users in its range, we will shortly say that $\mathcal B_l$ is empty. Then, given $u_i=q$ and $\mathcal M(q)= \mathcal I$, and given that $U_i$ is active, the probability that $U_i$ is collected can be expressed as
{\small{
\begin{align}
\label{eq-first}
& \mathbb P\left( U_i\,\mathrm{coll.}\,|\, u_i=q,\,\mathcal M(q)=\mathcal I,\,  U_i\mathrm{\;is\;active} \right)\nonumber = \mathbb P \left( \cup_{l \in \mathcal M(u_i)} \left\{ \mathcal B_l\mathrm{\;is\;empty}\right\}\,|\, u_i=q,\,\mathcal M(q)=\mathcal I,\,  U_i\mathrm{\;is\;active} \right)\nonumber \\
& = \mathbb P \left( \cup_{l \in \mathcal I} \left\{ \mathcal B_l \mathrm{\;is\;empty}\right\}\,|\, \mathcal M(q)=\mathcal I\right),
\end{align}}}
where in the last equality the two terms related with $U_i$ are dropped due to the fact that locations of base stations, and placements and activations of users different than $U_i$ are independent of the placement and activation of the user $U_i$.

Once the set of base stations in the range of the point $q$ is fixed, the event $\cup_{l\in \mathcal M(q)} \left\{ \mathcal B_l\mathrm{\;is\;empty}\right\}$ depends only on the positions of the base stations indexed in $\mathcal I$ and activation of users in the ranges of these base stations. In other words, this event is independent of the fact that, for any $k\notin \mathcal I$, the corresponding base station $B_k$ is placed outside the range of $q$. Noting that
 $
 \left\{ \mathcal M(q)=\mathcal I\right\}
 =\left\{b_l \in \mathbf B(q,r), l\in \mathcal I\right\}
 \cap
\left\{b_k \notin \mathbf B(q,r), k\notin \mathcal I\right\},
$
and combining this with the observation above, yields
\begin{align}
\label{eq-reduced-conditioning}
\mathbb P\left(\cup_{l\in \mathcal I} \left\{ \mathcal B_l\mathrm{\;is\;empty}\right\}\,|\,\mathcal M(q)=\mathcal I  \right)=\mathbb P\left(\cup_{l\in \mathcal I} \left\{ \mathcal B_l\mathrm{\;is\;empty}\right\}\,|\,
b_l\in \mathbf B(q,r),\,l\in \mathcal I  \right).
\end{align}

For $l=1,...,m$, denote by $F_l$ the event $\left\{b_l\in \mathbf B(q,r)\right\}$, and by $E_l$ the event $\left\{ \mathcal B_l\mathrm{\;is\;empty}\right\}$. To compute the right hand side in~\eqref{eq-reduced-conditioning}, we apply the inclusion-exclusion formula:
\begin{align}
\label{eq-inclusion-exclusion}
\mathbb P\left(\cup_{l\in \mathcal I} E_l \,|\, \cap_{l\in \mathcal I} F_l \right) =
\sum_{l_1 \in \mathcal I}\mathbb P \left( E_{l_1}\,|\, \cap_{l\in \mathcal I} F_l  \right)
\,-\, \sum_{(l_1,l_2) \in {\mathcal I \choose 2}} \mathbb P\left( E_{l_1}\cap E_{l_2}\,|\, \cap_{l\in \mathcal I} F_l \right) \,+\,\ldots\,\nonumber \\
\,+\, (-1)^{|\mathcal I|-1} \mathbb P\left( E_{l_1}\cap\,\ldots\,\cap E_{l_{\mathcal I}}\,|\, \cap_{l\in \mathcal I} F_l \right).
\end{align}
The first step in simplifying the preceding expression is to note that, for any fixed $k$-tuple $(l_1,...,l_k)$ of elements of $\mathcal I$ and any $l\in \mathcal I\setminus \left\{l_1,...,l_k\right\}$, the event $E_{l_1}\,\cap\,\ldots\,\cap E_{l_k}$ is independent of $F_l$. Since the independence holds for any such $l$, we have that $E_{l_1}\,\cap\,\ldots\,\cap E_{l_k}$ is independent of the intersection $\cap_{l\in \mathcal I \setminus \left\{l_1,...,l_k\right\}} F_l$. Thus,
 $
\mathbb P\left(  E_{l_1}\cap\,\ldots\,\cap E_{l_k}\,|\, \cap_{l\in \mathcal I} F_l\right)
$ $= \mathbb P\left(  E_{l_1}\cap\,\ldots\,\cap E_{l_k}\,|\,  F_{l_1}\cap\,\ldots\,\cap F_{l_k}\right),
 $
for any fixed $k=1,...,|\mathcal I|$, for any fixed $k$-tuple of elements of~$\mathcal I$.
Repeating this for each $k=1,...,|\mathcal I|$, and each $k$-tuple of elements of $\mathcal I$, from~\eqref{eq-inclusion-exclusion}:
\begin{align}
\label{eq-inclusion-exclusion-improved}
\mathbb P\left(\cup_{l\in \mathcal I} E_l \,|\, \cap_{l\in \mathcal I} F_l \right) =
\sum_{l_1 \in \mathcal I}\mathbb P \left( E_{l_1}\,|\, F_{l_1}  \right)
\,-\, \sum_{(l_1,l_2) \in {\mathcal I \choose 2}} \mathbb P\left( E_{l_1}\cap E_{l_2}\,|\, F_{l_1}\cap F_{l_2} \right) \,+\,\ldots\,\nonumber  \\
\,+\, (-1)^{|\mathcal I|-1} \mathbb P\left( E_{l_1}\cap\,\ldots\,\cap E_{l_{|\mathcal I|}}\,|\,
\, F_{l_1}\cap\,\ldots F_{l_{|\mathcal I|}} \right);
\end{align}
we note that, in the last term, $F_{l_1}\cap\,\ldots F_{l_{|\mathcal I|}}= \cap_{l\in \mathcal I} F_l$. We now focus on one term in the preceding sum that corresponds to a chosen $k$ and $(l_1,...,l_k)\in {\mathcal I \choose k}$. Put in simple terms, the event $E_{l_1}\cap\,\ldots\,\cap E_{l_k}$ means that there are no active users in any of the disks around base stations indexed in $\mathcal I$, which is equivalent to having no active users in the union of such disks. What we are then interested in is the probability of the latter event given that each of the base stations indexed in $\mathcal I$ lie not farther than $r$ from the given position $q$ of user $U_i$. Exploiting the symmetry of the base stations, we see that this probability is the same for any choice of $k$ different base stations, and hence for base stations $B_1,...,B_k$.
Therefore, for any $(l_1,...,l_k)\in {\mathcal I \choose k}$, and $\mathcal I\subseteq \{1,...,m\}$, we have,
 $
\mathbb P\left(  E_{l_1}\cap\,\ldots\,\cap E_{l_k}\,|\,  F_{l_1}\cap\,\ldots\,\cap F_{l_k}\right)=$ $
\mathbb P\left(  E_{1}\cap\,\ldots\,\cap E_{k}\,|\,  F_{1}\cap\,\ldots\,\cap F_{k}\right).
 $
Using the above identity for each of the terms in the sum in~\eqref{eq-inclusion-exclusion-improved}, and denoting with $d$ the cardinality of $\mathcal I$, yields
\begin{align}
\label{eq-inclusion-exclusion-twice-improved}
\mathbb P\left(\cup_{l\in \mathcal I} E_l \,|\, \cap_{l\in \mathcal I} F_l \right) =
d \,\mathbb P \left( E_{1}\,|\, F_{1}  \right)
\,-\, {d \choose 2} \mathbb P\left( E_{1}\cap E_{2}\,|\, F_{1}\cap F_{2} \right) \,+\,\ldots\, \nonumber \\
\,+\, (-1)^{d-1} \mathbb P\left( E_{1}\cap\,\ldots\,\cap E_{d}\,|\,  F_{1}\cap\,\ldots\,\cap F_{d} \right).
\end{align}
Remark that the probability in~\eqref{eq-inclusion-exclusion-twice-improved} depends on $\mathcal I$ only through its cardinality. Therefore,~\eqref{eq-inclusion-exclusion-twice-improved} holds not only for fixed $\mathcal I$ of cardinality $d$, but for all subsets of $\{1,...,m\}$ of the same cardinality.

We now compute $\mathbb P\left( E_{1}\cap\,\ldots\,\cap E_{k}\,|\, F_{1}\cap\,\ldots \cap F_{k}\right)$ for each fixed $k$, $1\leq k\leq m$ and for a given $q\in \mathcal A$ (recall that both $E_l$ and $F_l$ are defined with respect to a fixed location $q$ of the user $U_i$). To simplify the exposition, for $k=1,...,m$, we let:
 $
I_k(q):= \mathbb P\left( E_{1}\cap\,\ldots\,\cap E_{k}\,|\, F_{1}\cap\,\ldots F_{k}\right).
 $
 Suppose that base stations $B_1$,...,$B_k$ are placed, respectively, in $q_1$,...,$q_k$, where $q_l \in \mathbf B(q,r)$, $l=1,..,k$. Conditioned on $b_l=q_l$, $l=1,...,k$, the event $E_{1}\cap\,\ldots\,\cap E_{k}$ is equivalent to the event that there are no active users in the union $\cup_{l=1}^k \mathbf B(q_i,r)$ of the base stations' ranges. Note now that if $q\in \mathcal A^{\mathrm{o},r}$, then because each $q_l$ is within distance $r$ from $q$, we have that each of the balls $\mathbf B(q_l,r)$, $l=1,...,k$, belongs to $\mathcal A$, implying that the union $\cup_{l=1}^k \mathbf B(q_i,r)$ also belongs to $\mathcal A$. Let $\mathcal{U}(q_1,...,q_k)$ denote the area of $\cup_{l=1}^k \mathbf B(q_i,r)$. Now, a fixed user, say $U_j$, is not active in $\cup_{l=1}^k \mathbf B(q_i,r)$ if and only if: 1) $U_j$ either does not belong to $\cup_{l=1}^k \mathbf B(q_i,r)$;  or 2) $U_j$ belongs to $\cup_{l=1}^k \mathbf B(q_i,r)$, but it is inactive. Due to uniformity of the placements, the former happens with the probability equal to the area of $\mathcal A \setminus (\cup_{l=1}^k \mathbf B(q_i,r))$, which for $q\in \mathcal A^{\mathrm{o},r}$, equals $(1- \mathcal{U}(q_1,...,q_k))$. Similarly, for $q\in \mathcal A^{\mathrm{o},r}$, the latter happens with the probability equal to $\mathcal{U}(q_1,...,q_k)(1-1/\tau)$.
Summing up, we have that for any $q \in \mathcal A^{\mathrm{o},r}$, the probability that a fixed user is not active in $\cup_{l=1}^k \mathbf B(q_i,r)$ equals $\left( 1- {\mathcal U}(q_1,...,q_k)/\tau \right)$, and, by the independence among users:
\begin{equation}
\mathbb P\left( E_{1}\cap\,\ldots\,\cap E_{k}\,|\, F_{1}\cap\,\ldots F_{k}, b_l=q_l,\,l=1,...,k\right)
= (1- \mathcal U(q_1,...,q_k)/\tau)^{n-1},
\end{equation}
which holds for any fixed $q \in \mathcal A^{\mathrm{o},r}$ and $q_l\in \mathbf B(q,r)$, $l=1,...,k$. We now compute the joint conditional density of $b_1,...,b_k$ given that each $b_l$ belongs to $\mathbf B(q,r)$.
By the mutual independence of $b_l$'s, we have that, for any measurable set $D\subseteq \mathbb R^{2k}$,
 $
\mathbb P\left( (b_1,...,b_k) \in D \,|\,b_l\in \mathbf B(q,r),\,l=1,...,k \right) $ $=
\prod_{l=1}^k \mathbb P \left( b_l\in D_l\,|\, b_l\in \mathbf B(q,r) \right)
$ =$
\prod_{l=1}^k \left(\int_{(x_l,y_l) \in D_l} h_q(x_l,y_l)\,d x_l\, dy_l \right).
$
Here, $h_q(x,y)$ is the conditional density function of $b_l$ given that $b_l\in \mathbf B(q,r)$ (and it does not depend on $l$), and $D_l= \{(x_l,y_l)\in \mathbb R^2:$$ \, (x_1,y_1,...,x_l,y_l,...,x_k,y_k) \in D,$$ \mathrm{\;for\;some\;}x_j,y_j,...,j=1,...,k, j\neq l\}$, that is, $D_l$ is the projection of $D$ to the coordinates $l$ and $l+1$.
It is easy to show that, for any $l$, $h_q(x,y)$ is uniform:
 $
h_q( x,y )=
 \frac{1}{r^2 \pi },$ if $ (x,y)\in \mathbf B(q,r)$,
 and $h_q(x,y)=0$, else.
Returning to computing $I_k(q)$, summing up the previous conclusions yields:
{\small{
\begin{align}
I_k(q) = (r^2 \pi)^{-k}\int_{ (x_1,y_1) \in \mathbf B(q,r)}\cdots \int_{ (x_k,y_k) \in \mathbf B(q,r)}
  \, \left[\,1- \mathcal{U}( (x_1,y_1),...,(x_k,y_k))/\tau\,\right]^{n-1}\, d x_1\,d y_1\,\cdots\,d x_k\,dy_k.
\end{align}}}
Note that, as long as $q\in \mathcal A^{\mathrm{o},r}$, the value of $I_k (q)$ stays the same. We therefore drop the dependence on $q$ and simply write $I_k$ for $I_k(q)$ whenever $q\in \mathcal A^{\mathrm{o},r}$. Recall the variables $\alpha_k$'s and their distributions $\mu_k$'s
 in Section~\ref{section-performance-analysis}. Then, the integral $I_k$ can be written as:
\begin{equation}
\label{eqn-integral-k-proof}
I_k = \int_{a=1}^4 (1-r^2\pi a/\tau)^{n-1} d\mu_k(a).
\end{equation}

Combining now~\eqref{eq-first},~\eqref{eq-reduced-conditioning}, and ~\eqref{eq-inclusion-exclusion-twice-improved}, we obtain that for any $q\in \mathcal A^{\mathrm{o},r}$, and any $\mathcal I\subseteq \{1,...,m\}$
{\small{
\begin{align}
\mathbb P\left( U_i\,\mathrm{coll.}\,|\, u_i=q,\,\mathcal M(q)=\mathcal I,\,  U_i\mathrm{\;is\;active} \right)\nonumber  =  d {{I_1}}  \,-\, {d \choose 2} {{I_2}} \,+\,\ldots\, \,+\, (-1)^{k-1} {d \choose k} I_k
\,+\,\ldots\,\,+\, (-1)^{d-1} {{I_d}},
\end{align}}}
where, we recall, $d= |\mathcal I|$.
Summing up over different $\mathcal I$, and using the fact that event $\{\mathcal M(q) = \mathcal I\}$ is independent of the position and activation of user $U_i$,
{\small{
\begin{align}
\label{eq-almost-final}
&\mathbb P\left( U_i\,\mathrm{coll.}\,|\, u_i=q,\,U_i\mathrm{\;is\;active} \right) =  \sum_{\mathcal I \subseteq \{1,...,m\}, \mathcal I\neq \emptyset}
\mathbb P\left( U_i\,\mathrm{coll.}\,|\, u_i=q,\,\mathcal M(q)=\mathcal I,\,  U_i\mathrm{\;is\;active} \right)
\mathbb P\left(\mathcal M(q)=\mathcal I \right)\nonumber \\
&=\sum_{d=1}^m\left( d {{I_1}} \,-\, {d \choose 2} {{I_2}} \,+\,\ldots\, 
\,+\,\ldots\,\,+\, (-1)^{d-1} {{I_d}} \right)   {m \choose d}   (r^2 \pi)^{d} (1-r^2 \pi)^{m-d}.
\end{align}}}
For each $k=1,...,m$, sum up in ${\zeta}_k$ all the terms that multiply $I_k$,
\begin{equation}
\label{eqn-lambda-k}
{\zeta}_k= \sum_{d=k}^m\,{d \choose k} \,{m \choose d}   (r^2 \pi)^{d} (1-r^2 \pi)^{m-d}=\sum_{d=k}^m\,{d \choose k}\Delta_d.
\end{equation}
We can then compactly write~\eqref{eq-almost-final} as
\begin{align}
\mathbb P\left( U_i\,\mathrm{coll.}\,|\, u_i=q,\,U_i\mathrm{\;is\;active} \right)= {\zeta}_1 {{I_1}}  - {\zeta}_2  {{I_2}} +...+ (-1)^{m-1} {\zeta}_m I_m,
\end{align}
where the $I_k$'s are given in~\eqref{eqn-integral-k-proof}.
Note that the obtained identity holds for all $q\in \mathcal A^{\mathrm{o},r}$.
To finalize the analysis, it only remains to integrate over different $q\in \mathcal A$. We split the integration to $q\in \mathcal A^{\mathrm{o},r}$ and $q\in \partial \mathcal A^{r}$,
 $
\mathbb P\left( U_i\,\mathrm{coll.}\,|\,U_i\mathrm{\;is\;active} \right) $ $=
\mathbb P\left( U_i\,\mathrm{coll.}\,|\,u_i \in \mathcal A^{\mathrm{o},r},\, U_i\mathrm{\;is\;active} \right) (1-4r)^2 $
 $+ \mathbb P\left( U_i\,\mathrm{coll.}\,|\,u_i \in \partial \mathcal A^{r},\, U_i\mathrm{\;is\;active} \right)  (1-(1-4r)^2).
 $
As~$\mathbb P\left( U_i\,\mathrm{coll.}\,|\,u_i \in \partial \mathcal A^{r},\, U_i\mathrm{\;is\;active} \right) \in [0,1]$,
we finally obtain the upper and lower bounds in Theorem~\ref{theorem-non-coop}~(a).
%
%


\textbf{Proof of Theorem~1, part~(b)}.
We now consider the asymptotic setting. Note that,
as $r \rightarrow 0$ in the asymptotic setting, the left and right inequalities in Theorem~\ref{theorem-non-coop}, part~(a)
 both converge to the limit of $P_{\mathrm{coll.}}^{\mathrm{o},r}$. Therefore, it remains to find the limit of~$P_{\mathrm{coll.}}^{\mathrm{o},r}$.

We first show that $I_k $ converges to
$I_{\infty,k}:=\int_{a=1}^4 e^{-\delta G a}d\mu_k(a)$ in~\eqref{eqn-asymptotic-non-coop}.
 First, note that the function:
\[
\phi_n(a)=(1-r^2\pi \,a/\tau)^{n-1} \rightarrow e^{-\delta\,G \, a }, \:\forall a \in [1,4].
\]
This is because $\frac{(n-1)r^2\pi}{\tau} = (m r^2 \pi)\frac{(n-1)}{\tau\,m} \rightarrow \delta\,G$.
 Denote now $\epsilon_n(a)=|e^{-\delta\,G a } -\phi_n(a)|$,
and by $\epsilon_n^\star:=\sup_{a \in [1,4]}\epsilon_n(a)$.
 Note that:
 \begin{eqnarray*}
 \epsilon_n(a) &=& | e^{-{\delta G} a} - e^{-(n-1)ar^2 \pi /\tau} + e^{-(n-1)ar^2 \pi /\tau} - (1-a r^2 \pi /\tau)^{n-1}|\\
 &\leq&
 | e^{-{\delta G} a} - e^{-(n-1)ar^2 \pi /\tau}| + |e^{-(n-1)ar^2 \pi /\tau} - (1-a r^2 \pi /\tau)^{n-1}|\\
 &=&
 e^{-{\delta G} a}|1-e^{-a({\delta G}-(n-1)r^2\pi /\tau)}| + |e^{-(n-1)ar^2 \pi /\tau} - (1-a r^2 \pi /\tau)^{n-1}|,
 \end{eqnarray*}
 and so
 {\small{
 \begin{eqnarray*}
 \epsilon_n^\star \leq e^{-{\delta G}} \max
 \left\{|1-e^{-4|{\delta G}-(n-1)r^2 \pi /\tau|}|,\, |1-e^{|({\delta G}-(n-1)r^2 \pi /\tau)|}|\right\}
 + |e^{-4(n-1)r^2 \pi /\tau}-(1-4 r^2\pi /\tau)^{n-1} |,
 \end{eqnarray*}}}
  which
 converges to zero as $n \rightarrow \infty$. Therefore:
 \begin{eqnarray}
 \left|I_k -\int_{1}^4  e^{-{\delta G}\, a } d \mu_k(a) \right|
 \leq \int_1^4 \epsilon_n(a) d \mu_k(a) \leq \epsilon_n^\star \rightarrow 0,
 \end{eqnarray}
 and so:
 \begin{equation}
 \label{eqn-S-k-asymptotic}
 I_k \rightarrow I_{\infty,k}:=\int_1^4 e^{-{\delta G} a} d \mu_k(a)\:\:\mathrm{as \,}n\rightarrow \infty,\,\forall k.
 \end{equation}

Next, we show that the quantity $\zeta_k$ in~\eqref{eqn-P-zeta-k} converges to $ \delta^k/k!$.
 Consider the term $\Delta_d$--the probability that a binomial random variable
  with parameters $m$ (number of trials) and $r^2 \pi$ (success probability) equals $d$.
    It is well known that, when $m \rightarrow \infty$, $r^2 \pi \rightarrow 0$,
    and $m r^2 \pi \rightarrow {{\delta}}$, ${{\delta}}>0$ the binomial distribution
    converges to the Poisson distribution with parameter ${{\delta}}$; that is,
    for all $d$, $ {m \choose d}(r^2\pi)^d(1-r^2\pi)^{m-d}$ converges to
    $e^{-{{\delta}}}{{\delta}}^d/d !$. Therefore, when $n \rightarrow \infty$, $\zeta_k$ converges to:
    \[
    \sum_{d=k}^{\infty} {d \choose k} e^{-{{\delta}}}\frac{{{\delta}}^d}{d!}.
    \]
    We further simplify the resulting expression and obtain the desired result as follows:
    \begin{eqnarray*}
    &\,& e^{-{{\delta}}}\frac{{{\delta}}^k}{k!}
    \sum_{d=k}^{\infty} \frac{d!}{(d-k)!}\frac{{{\delta}}^{d-k}}{d!}\\
    &=&
    e^{-{{\delta}}}\frac{{{\delta}}^k}{k!}
    \sum_{d=k}^{\infty} \frac{{{\delta}}^{d-k}}{(d-k)!} = \frac{{{\delta}}^k}{k!}.
    \end{eqnarray*}
Applying the established facts that $I_k\rightarrow \int_{a=1}^4 e^{-\delta G a}d\mu_k(a)$ and
$\zeta_k \rightarrow \delta^k/k!$, and using the fact
that $r \rightarrow 0$, we finally obtain the desired result.

It remains to prove the lower bound in~\eqref{eqn-asymptotic-non-coop}.
We do this by relying on the proof of part~(a).
Consider~$\mathbb P\left( U_i\,\mathrm{coll.}\,|\, u_i=q,\,\mathcal M(q)=\mathcal I,\,  U_i\mathrm{\;is\;active} \right)$
 $=
\mathbb P\left(\cup_{l\in \mathcal I} E_l \,|\, \cap_{l\in \mathcal I} F_l \right)$, for
a fixed
$\emptyset \neq \mathcal M(q) \subset \{1,...,m\}$.
Note that $\mathbb P\left(\cup_{l\in \mathcal I} E_l \,|\, \cap_{l\in \mathcal I} F_l \right)
 \geq \mathbb P\left(E_{l_1} \,|\, \cap_{l\in \mathcal I} F_l \right)$ (where $l_1$
  is an arbitrary index in $\mathcal I$),
 which, as shown in the proof of part~(a),
 equals~$\mathbb P\left(E_{l_1} \,|\, F_{l_1} \right)$,
 and further equals $I_1(q)
 =(1-r^2\pi/\tau)^{n-1}$. Summing over all the $\mathcal I$'s
 different than empty set, as in~\eqref{eq-almost-final}, we obtain:
$\mathbb P\left( U_i\,\mathrm{coll.}\,|\, u_i=q,\,U_i\mathrm{\;is\;active} \right) $
$\geq (1-r^2\pi/\tau)^{n-1} (\,1-\mathbb P(\mathcal M(q)=\emptyset)\,)
$ $= (1-r^2\pi/\tau)^{n-1} (1-(1-r^2\pi)^m)$. Integrating
over all nominal placements, and passing
to the asymptotic setting, the result follows. This completes the proof of Theorem~\ref{theorem-non-coop}.

\subsection{Proof of Lemma~\ref{lemma-spatial-coop}}
%
%
%
%
Fix a user~$U_i$, and suppose it is active and has an arbitrary nominal placement~$q$.
We next lower bound $\mathbb P \left(U_i\,\mathrm{coll.}\,|\,u_i=q\right)$.
 Consider the following two events: $\mathcal{E}_1$--$U_i$ has no adjacent base stations; and
 $\mathcal{E}_2$--there exists at least one base station in $\mathbf{B}(q,r/2)$,
 there exists at least one user $U_j$, $j \neq i$, in $\mathbf{B}(q,r/2)$,
 and there are no base stations in $\mathbf{R}(q,r/2,3r/2)$.
  The events $\mathcal{E}_1$
  and $\mathcal{E}_2$ are disjoint. Further, clearly, $U_i$
  is not collected if $\mathcal{E}_1$ occurs. It is not difficult to see that
  $U_i$ is not collected if $\mathcal{E}_2$ occurs, also. Namely, if $\mathcal{E}_2$ occurs,
  $U_i$ is located in a complete bipartite graph $\mathcal{G}_2$, a subgraph of $\mathcal G_0$. The graph $\mathcal{G}_2$ contains
  $n_2 \geq 2$ users (precisely those lying in $\mathbf{B}(q,r/2)$),
  and $m_2\geq 1$ base stations (those lying in $\mathbf{B}(q,r/2)$). The base stations in
  $\mathcal{G}_2$ may be connected to users outside $\mathcal{G}_2$, but
  the users in $\mathcal{G}_2$ are not connected to other base stations. This is ensured
  by having no base stations in~$\mathbf{R}(q,r/2,3r/2)$. (See the Supplementary material for an illustration of $\mathcal G_2$.) Then, all the base stations
  adjacent to $U_i$ have at least two neighboring users from $\mathcal{G}_2$ and are ``blocked.''
   In other words, the set of users that belong to $\mathcal{G}_2$ is a stopping set.
  Hence, $U_i$ is not collected if $\mathcal{E}_2$ occurs. Summarizing:
\begin{eqnarray}
\label{eqn-equation-appendix}
\mathbb P (U_i\,\mathrm{not\,coll.}\,|\,u_i=q, \,U_i\,\mathrm{act.}) &\geq& \mathbb P\left(\mathcal{E}_1\cup \mathcal{E}_2\,|\,u_i=q
, \,U_i\,\mathrm{act.}\right)\\
&=&
\mathbb P\left(\mathcal{E}_1\,|\,u_i=q, \,U_i\,\mathrm{act.}\right)+\mathbb P\left(\mathcal{E}_2\,|\,u_i=q
, \,U_i\,\mathrm{act.}\right)=:p_1+p_2, \nonumber
\end{eqnarray}
where the second from last equality holds because $\mathcal{E}_1$ and $\mathcal{E}_2$ are disjoint.
We now evaluate $p_1$ and $p_2$. We have that $p_1=\left( 1-r^2\pi\right)^{m}$, which converges
asymptotically to~$\mathrm{exp}(-\delta)$. For $p_2$, we have:
 $
p_2=\left[1-\left( 1-\frac{pr^2 \pi}{4}\right)^{n-1} \right] $$ \left[ 1-\left( 1-\frac{r^2 \pi}{4(1-2r^2\pi)}\right)^{m}\right]$ $\left[ 1-2 r^2 \pi\right]^m.
 $
The first term above is the probability of having at least one user $U_j$, $j \neq i$, in $\mathbf{B}(q,r/2)$.
The second term is the probability of having
at least one station in $\mathbf{B}(q,r/2)$, conditioned on having no stations in $\mathbf{R}(q,r/2,3r/2)$.
The third term is the probability of having no stations in $\mathbf{R}(q,r/2,3r/2)$.
Asymptotically, $p_2$ converges to $(1-\mathrm{exp}(-\delta G/4))\,(1-\mathrm{exp}(-\delta/2))\,\mathrm{exp}(-2\delta)$.
 Applying the above results for $p_1$ and $p_2$ in~\eqref{eqn-equation-appendix}, and passing to the limit (where boundary effects vanish), we obtain the desired result.

\subsection{Proof of Theorem~\ref{theorem-temporal-coop-decod-prob}} Fix an arbitrary user $U_i$ at arbitrary nominal placement $q\in \mathcal{A}^{o,r}$. Because $r\rightarrow 0$ as $n \rightarrow \infty$, it suffices to lower bound
$\mathbb P(U_i\,\mathrm{coll.}\,|\,u_i=q)$ for any $q \in \mathcal{A}^{o,r}$. (We strictly show why this is sufficient later in the proof.)
 Denote by $N_B(u_i)$ the number of base stations in $\mathbf{B}(u_i,r)$,
and by $N_U(u_i)$ the number of users different than $U_i$ in $\mathbf{B}(u_i,2r)$.
  We first explain the intuition behind the proof, and then we formalize it through equations.
 We construct a specific scenario when $U_i$ is collected and evaluate its probability.
 The scenario is as follows: user $U_i$ has at least one base station in its $r$-neighborhood ($N_B(u_i) \geq 1$),
 and there are at most $C$ users different than $U_i$ in the $U_i$'s $2r$-neighborhood ($N_U(u_i) \leq C$).
 Without loss of generality, let $B_1$ be one of the base stations in $\mathbf{B}(u_i,r)$. In the considered scenario, $B_1$
  has in its neighborhood at most $C+1$ users. Then, the probability that $U_i$ is collected is greater than or equal
  the probability that $U_i$ is collected by $B_1$ working as a single base station system (in the sense of the system described in
   Subsection~\ref{subsection-single-base-stations})
   with $C+1$ users, i.e., with load $H=(C+1)/\tau$.

We now proceed with formalizing the above idea. We have:
{\small{
\begin{eqnarray*}
&\,&\mathbb P(U_i\,\mathrm{coll.}\,|\,u_i=q) \\
&\geq& \mathbb P(U_i\,\mathrm{coll.}\,|\,N_B(u_i) \geq 1,\,N_U(u_i)\leq C,\,u_i=q)\\
&\times& \mathbb P \left(  N_B(u_i) \geq 1,\,N_U(u_i) \leq C\,|\,u_i=q       \right).
\end{eqnarray*}}}
Next, note that:
\begin{eqnarray*}
&\,& \mathbb P \left(  N_B(u_i) \geq 1,\,N_U(u_i) \leq C\,|\,u_i=q       \right)\\
&=& \mathbb P \left(  N_B(q) \geq 1,\,N_U(q) \leq C\,|\,u_i=q       \right) \\
&=&\mathbb P \left(  N_B(q) \geq 1\right) \mathbb P \left(  N_U(q) \leq C\right),
\end{eqnarray*}
where the last equality holds by the independence of the users' and base stations' placements.
 Denote by $ \mathbb P(U_i\,\mathrm{coll.}\,|\,N_B(u_i) \geq 1,\,N_U(u_i)\leq C,\,u_i=q) = \widehat{P}$.
  We have:
\begin{eqnarray}
\label{eqn-proof-1}
\mathbb P(U_i\,\mathrm{coll.}\,|\,u_i=q)
\geq \widehat{P}\,\mathbb P \left(  N_B(q) \geq 1\right) \,\mathbb P \left(  N_U(q) \leq C\right).
\end{eqnarray}
Note that $N_B(q)$ is a binomial random variable with the number of trials
equal $m$ and success probability $r^2\pi$. Similarly, $N_U(q)$ is a binomial random variable with the number of trials
equal $n-1$ and success probability $4 r^2\pi$, and $\mathbb E[N_U(q)]=4 r^2\pi (n-1)$. From now on, we set
$C=\theta 4 r^2 \pi(n-1)$, for some $\theta >0$ that we specify later. We proceed by separately lower bounding each of the three probabilities on the right hand
side of~\eqref{eqn-proof-1}.

\textbf{Lower bound on $\widehat{P}$}. As explained in the intuition above, we have $\widehat P \geq P_{\mathrm{single}}$,
 where $P_{\mathrm{single}}$ is the probability that a fixed user $U_i$ is collected by the single base station system
 with $C+1=\theta(n-1)4 r^2\pi+1$ users, users' degree distribution $\Lambda$, and load
 $H=(C+1)/\tau$. (The term $1$ in $C+1$ comes from the inclusion of $U_i$ as well.) Note that we use here
 the fact that decoding probability with the single base station system
 is a monotonically non-increasing function of load $H$. (Conditioned on
 the number of served users be at most $C+1$, the worst case occurs for the number of users equal~$C+1$.)
 Next, note that $H=(C+1)/\tau = \frac{4 \theta (n-1) r^2\pi +1}{\tau}\frac{m}{m}
 =4 \delta G + o(1)$. Thus, we conclude that $\widehat{P}$ is asymptotically
 lower bounded by:
 \begin{equation}
 \label{eqn-proof-combine-1}
 {\rho}(H=4 \theta \delta G),
 \end{equation}
 where we recall that ${\rho}(H)$
  is the asymptotic decoding probability of the single base station system under load~$H$.
 %
 %
 %
 %
 %
 %

\textbf{Lower bound on $\mathbb P(N_B(q) \geq 1)$}. Clearly, $\mathbb P(N_B(q) \geq 1)=\mathbb P(U_i\,\mathrm{cov.}\,|\,u_i=q)$, and hence:
 \begin{equation}
  \label{eqn-proof-combine-2}
  \mathbb P(N_B(q) \geq 1) \rightarrow 1-e^{-\delta}.
  \end{equation}

\textbf{Lower bound on $\mathbb P(N_U(q) \leq \theta 4(n-1)r^2 \pi)$}.
 We use the Chebyshev inequality for the Binomial random variable $\zeta$ with
 the number of trials $\nu$ and success probability $\pi$; for any $\epsilon>0$:
 \begin{eqnarray*}
 \mathbb P(\,\zeta \geq (1+\epsilon) \mathbb E[\zeta]\,) \leq
 \mathbb P(\,|\zeta-\mathbb E[\zeta]| \geq \epsilon \mathbb E[\zeta]\,)
 \leq \frac{\mathrm{Var}(\zeta)}{\epsilon^2 (\mathbb E[\zeta])^2}
 = \frac{1-\pi}{\epsilon^2 \nu \pi},
 \end{eqnarray*}
 where the equality follows by replacing
 $\mathbb E[\zeta]=\nu \pi$, and $\mathrm{Var}(\zeta)=\nu \pi (1-\pi)$.
   Applying the above inequality to $N_U(q)$, with $\theta:=1+\epsilon$:
%
%
%
%
%
%
%
\begin{equation*}
\mathbb P(N_U(q) \leq \theta 4(n-1)r^2 \pi) \geq 1 - \frac{1-4 r^2 \pi}{\epsilon^2 (4 r^2 \pi)(n-1)} .
\end{equation*}
Note that $(n-1)r^2\pi = n r^2 \pi - r^2 \pi
= n r^2 \pi \frac{\tau m}{\tau m} - r^2 \pi = G \delta \tau - r^2\pi \rightarrow +\infty
$, as $n\rightarrow \infty$ (because $\tau(n) \rightarrow \infty.$) Thus, we conclude that:
\begin{equation}
\label{eqn-combine-3}
 \mathbb P(N_U(q) \leq \theta 4(n-1)r^2 \pi) \rightarrow 1, \,\,\theta=1+\epsilon,\,\,\,\mathrm{for\,all}\,\,
\epsilon>0.
\end{equation}
Now, we combine~\eqref{eqn-proof-combine-1}, \eqref{eqn-proof-combine-2}, and~\eqref{eqn-combine-3}, with $\theta=1+\epsilon$; we obtain
 that, $\forall q \in \mathcal A^{o,r}$, $\mathbb P(U_i\,\mathrm{coll.}\,|\,u_i=q)$ is asymptotically lower bounded by:
 \begin{equation}
 \label{eqn-asympt-lower-bound}
 {\rho}(H=(1+\epsilon)4\,\delta\, G)(1-e^{-\delta}),\:\:\forall \epsilon>0.
 \end{equation}
  %
 %
 Finally, note that $\mathbb P(u_i \in \mathcal{A}^{o,r} ) = (1- 4 r)^2$,
 which converges to one. Also,
 $\mathbb P(U_i\,\mathrm{coll.}) \geq \mathbb P(U_i\,\mathrm{coll.}\,|\,u_i  \in \mathcal{A}^{o,r})
 \mathbb P(u_i \in \mathcal{A}^{o,r} )$. Combining the last two observations,
 we finally obtain the desired result.

\subsection{An intuition for the and-or-tree heuristic with spatio-temporal cooperation}
\begin{figure}[thpb]
      \centering
      \vspace{-7mm}
      \includegraphics[height=2.3 in,width=3.3in]{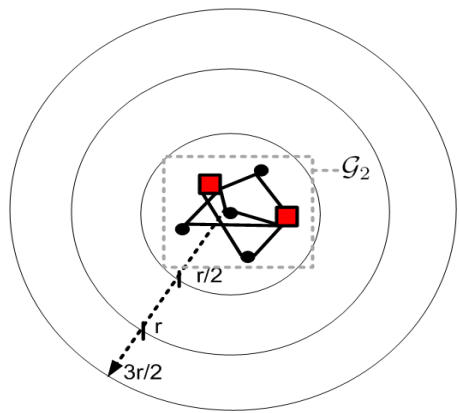}
       \vspace{-5mm}
       \caption{Illustration for the proof of Lemma~\ref{lemma-spatial-coop}. Graph $\mathcal G_2$ contains $m_2=2$ base stations and
       $n_2=4$ users.}
       \label{figure-appendix-1}
       \vspace{-7mm}
\end{figure}

\begin{figure}[thpb]
      \centering
      \vspace{-5mm}
       \includegraphics[height=2. in,width=2.6 in]{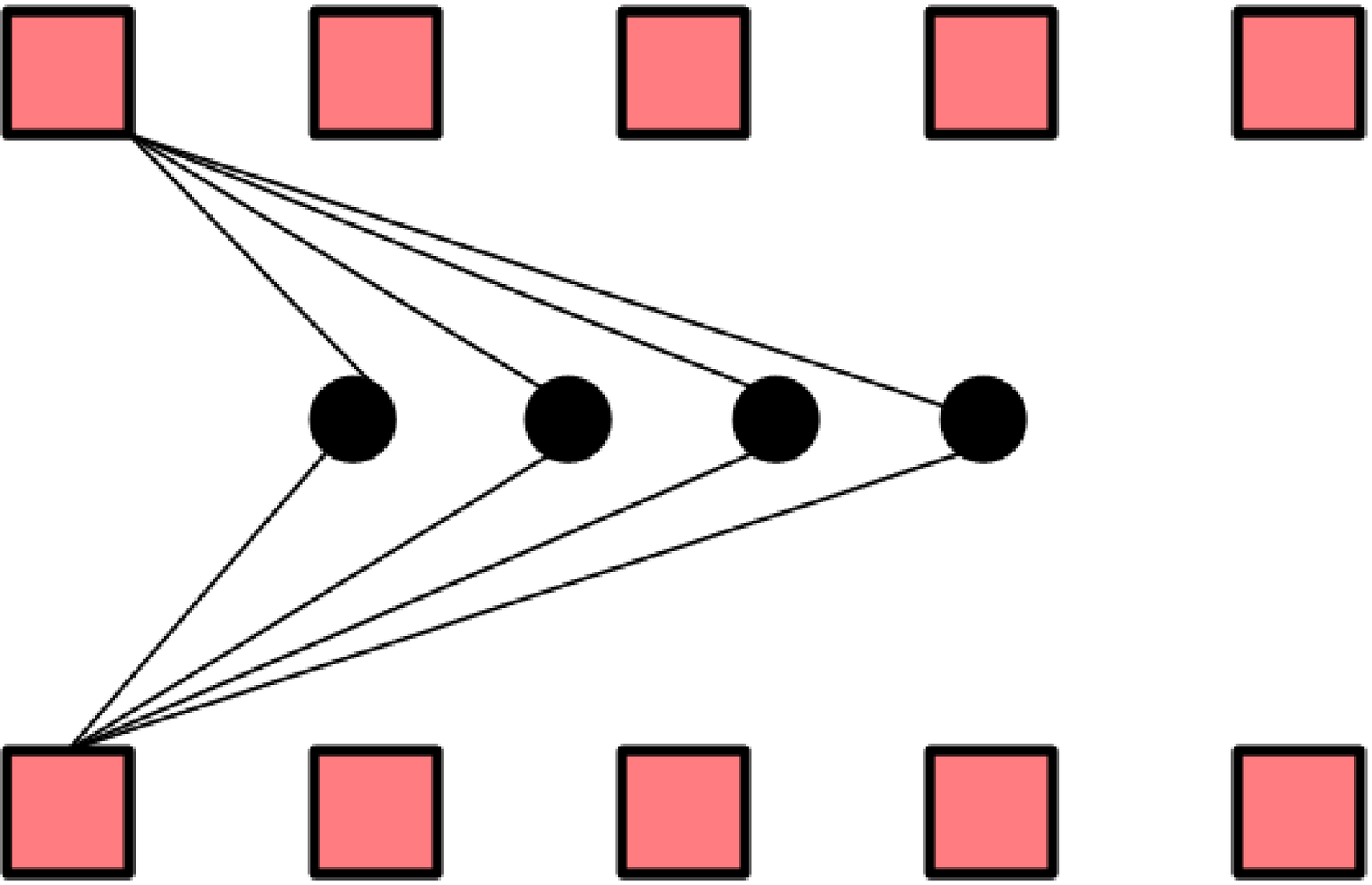}
       \includegraphics[height=1.9 in,width=2.5 in]{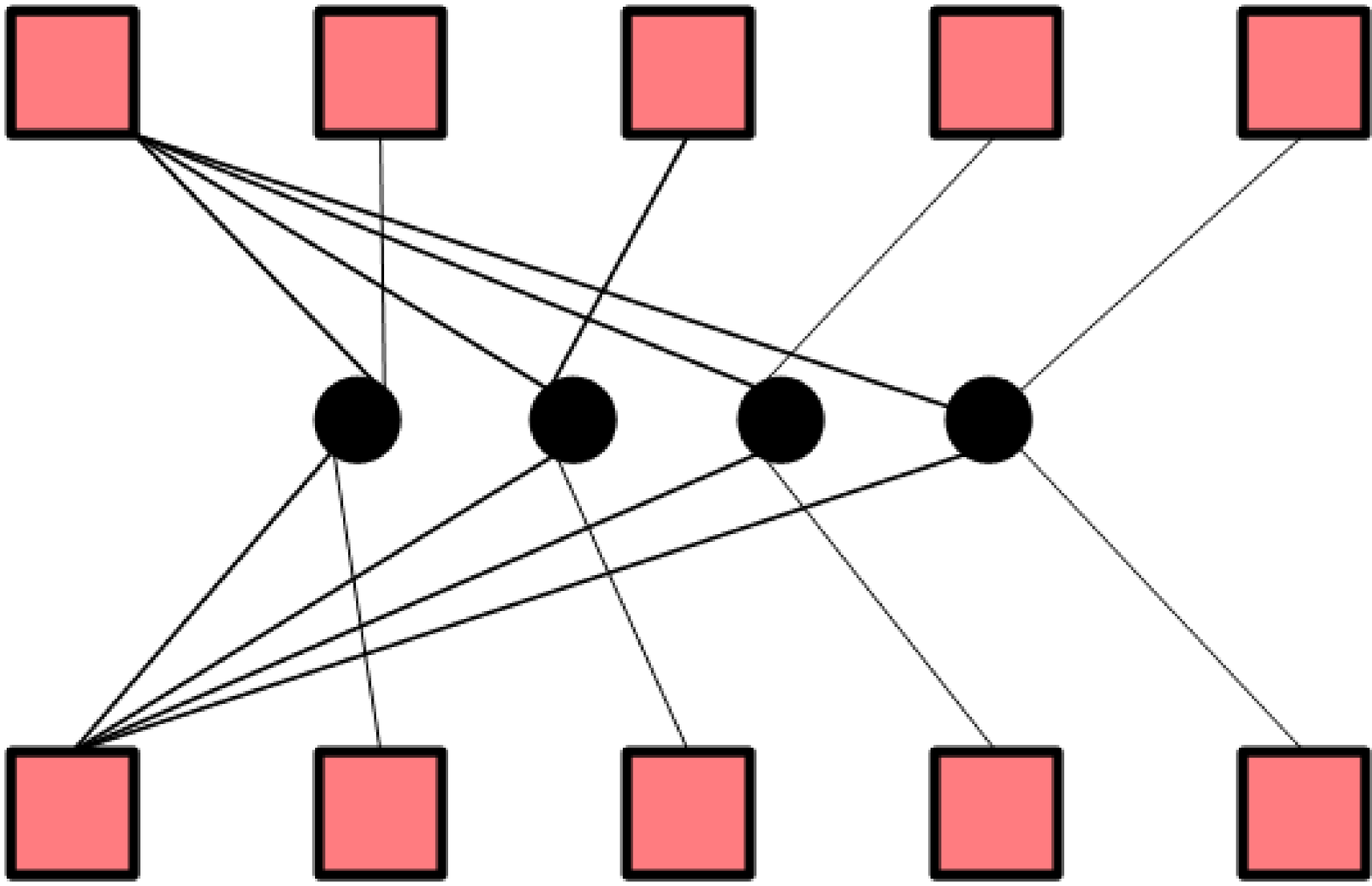}
       \includegraphics[height=1.9 in,width=2.5 in]{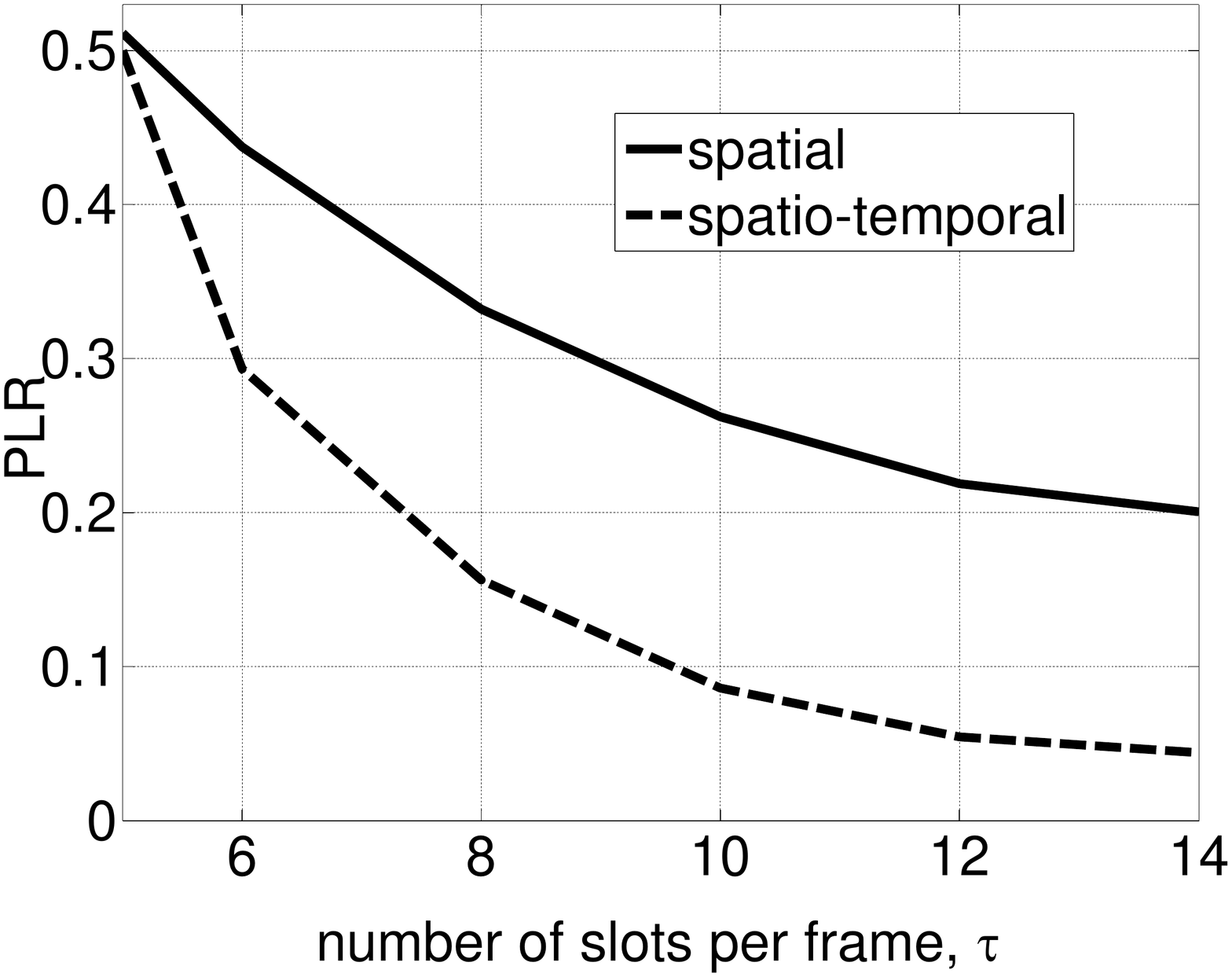}
       \vspace{-5mm}
       \caption{Top: Illustration for an intuition for the and-or-tree heuristic with spatio-temporal cooperation.
       The system is the system shown in Figure~\ref{figure-appendix-1}.
       There are $\tau=5$ slots, $2$ base stations, and $4$ users.
       The check nodes that correspond to base station $B_1$ are represented above users,
       while the check nodes that correspond to base station $B_2$ are represented below users.
       The top left Figure presents a scenario with spatial cooperation, while the top right Figure
       presents a corresponding scenario with spatio-temporal cooperation.
       While spatial cooperation collects no users, spatio-temporal cooperation collects all the four users.
        Bottom: PLR versus number of slots per frame $\tau$
        for the system in Figure~\ref{figure-appendix-1}. The solid (respectively, dashed) line corresponds to
        spatial (respectively, spatio-temporal) cooperation.}
       \label{figure-appendix-2}
       \vspace{-5mm}
\end{figure}

 We noted that, with spatial cooperation, and-or-tree heuristic may give over-optimistic
 performance estimates due to the emergence of local stopping sets.
 A major impact is played by the local stopping sets explained in Lemma~\ref{lemma-spatial-coop}.
 We provide here an intuitive explanation why
 the effect of the local stopping sets is reduced with spatio-temporal cooperation,
 thus leading to better predictions via and-or-tree evaluation (in the range of the system
  parameters of interest). Consider user $U_i$ at location $q$ (the user at the center of the circles in Figure~\ref{figure-appendix-1}),
  suppose that there are 2 base stations and 4 users in $\mathbf{B}(q,r/2)$, and no
  base stations in $\mathbf{R}(q,r/2,3r/2)$. Also, for simplicity,
  suppose there are no users in $\mathbf{R}(q,r/2,3r/2)$ (although the last condition is
  not imposed in the proof of Lemma~\ref{lemma-spatial-coop}.) The corresponding
  system is illustrated in Figure~\ref{figure-appendix-1}.
   Note that the users and base stations in Figure~\ref{figure-appendix-1} are isolated from
  the rest of the system. Now, consider spatial cooperation. Suppose that there are~$\tau=5$ slots and that each of the four
  users transmits at slot~$1$. This is illustrated in Figure~\ref{figure-appendix-2} (left).
  In this case, all the users are ``blocked'' and none of them is collected.
  Hence, the local stopping set disables decoding of the users.
  Now, consider spatio-temporal cooperation where each user transmits according to
  the constant-degree-two distribution. Suppose again that each of the four
  users transmitted at slot~$1$. While this scenario disables decoding with spatial cooperation,
  spatio-temporal cooperation still allows the decoding of all (or a subset of) users with a certain probability.
   One successful scenario is depicted in Figure~\ref{figure-appendix-2}~(right).
   To be concrete, we plot in Figure~\ref{figure-appendix-2} (bottom)
    a Monte Carlo estimate of PLR (probability that a fixed user is not collected)
    versus $\tau$ for the system in Figure~\ref{figure-appendix-1}
     with $2$ base stations and $4$ users. (Here, when calculating PLR,
     we average over the user activations--slot selections.) We can clearly see that
     the ``blocking''
      geometric structure as in Figure~\ref{figure-appendix-1} affects much more spatial cooperation than spatio temporal cooperation.

\end{document}